\documentclass{article}[dvipsnames, 11pt]
\usepackage[a4paper, margin=1.5in]{geometry}
\usepackage[utf8]{inputenc}
\usepackage{hyperref}
\usepackage{xcolor}
\usepackage{float}
\usepackage{graphicx}
\usepackage{titlesec}
\usepackage{placeins}
\usepackage{enumitem}
\usepackage{csquotes}
\usepackage{amsfonts}
\usepackage{amsmath}
\usepackage{amssymb}
\usepackage{bbm}
\usepackage{enumitem}
\usepackage{booktabs}
\usepackage[range-phrase={\text{\ensuremath{-}}}]{siunitx}
\usepackage{subfig}
\usepackage{colortbl}
\usepackage{tikz}
\usepackage{tablefootnote}
\usepackage{makecell}
\usepackage{hyperref}
\usepackage[resetlabels]{multibib}
\newcites{AC}{References}
\newcites{AD}{References}
\newcites{AE}{References}
\usepackage{authblk}

\usepackage{array}
\usepackage{multirow}

\usepackage{longtable}
\usetikzlibrary{shapes.geometric, arrows, fit}

\titleformat{\paragraph}
{\normalfont\normalsize\bfseries}{\theparagraph}{1em}{}
\titlespacing*{\paragraph}
{0pt}{3.25ex plus 1ex minus .2ex}{1.5ex plus .2ex}

\newcolumntype{L}[1]{>{\raggedright\let\newline\\\arraybackslash\hspace{0pt}}m{#1}}
\newcolumntype{C}[1]{>{\centering\let\newline\\\arraybackslash\hspace{0pt}}m{#1}}
\newcolumntype{R}[1]{>{\raggedleft\let\newline\\\arraybackslash\hspace{0pt}}m{#1}}
\usepackage[textsize=footnotesize]{todonotes}

\newcommand{\WO}{ETKAS}
\newcommand{\WOP}[0]{\WO\ point system}

\usepackage{setspace}
\usepackage{array}

\def\Plus{\texttt{+}}

\title{The ETKidney simulator: a discrete event simulator to assess the impact of alternative kidney allocation rules in Eurotransplant}
\author[a,b]{H.C. de Ferrante\thanks{Corresponding author. Emails: h.c.d.ferrante@tue.nl; h.deferrante@eurotransplant.org}}
\author[b]{R. Laguna Goya}
\author[a]{B.M.L. Smeulders}
\author[a]{F.C.R. Spieksma}
\author[b]{I. Tieken}
\affil[a]{Department of Mathematics and Computer Science, Eindhoven University of Technology (Postal address: De Groene Loper 5, Eindhoven, the Netherlands)}
\affil[b]{Eurotransplant International Foundation (Postal address: Haagse Schouwweg 6, Leiden, the Netherlands)}

\date{February 2025}

\date{February 2025}

\begin{document}

\maketitle

\begin{abstract}	
	Over 10,000 candidates wait for a kidney transplantation in Eurotransplant, and are prioritized for transplantation based on the allocation rules of the Eurotransplant Kidney Allocation System (ETKAS) and Eurotransplant Senior Program (ESP). These allocation rules have not changed much since ETKAS and ESP's introductions in 1996 and 1999, respectively, despite identification of several areas of improvement by the Eurotransplant Kidney Advisory Committee (ETKAC). A barrier to modernizing ETKAS and ESP kidney allocation rules is that Eurotransplant lacks tools to quantitatively assess the impact of policy changes.
	\par
	We present the ETKidney simulator, which was developed for this purpose. This tool simulates kidney allocation according to the actual ETKAS and ESP allocation rules, and implements Eurotransplant-specific allocation mechanisms such as the system which is used to balance the international exchange of kidneys. The ETKidney simulator was developed in close collaboration with medical doctors from Eurotransplant, and was presented to ETKAC and other major stakeholders. To enhance trust in the ETKidney simulator, the simulator is made publicly available with synthetic data, and is validated by comparing simulated to actual ETKAS and ESP outcomes between 2021 and 2024.	
	\par
	We also illustrate how the simulator can contribute to kidney policy evaluation with three clinically motivated case studies. We anticipate that the ETKidney simulator will be pivotal in modernizing ETKAS and ESP allocation rules by enabling informed decision-making on kidney allocation rules together with national competent authorities.
\end{abstract}

\newpage
\section{Introduction}
Over 10,000 patients wait for a kidney transplantation in the countries served by Eurotransplant (Austria, Belgium, Croatia, Germany, Hungary, Luxembourg, the Netherlands, and Slovenia). Eurotransplant is responsible for the allocation of kidneys from deceased donors to these patients, and places most kidneys through two programs: the Eurotransplant Kidney Allocation System (ETKAS) for kidneys from donors aged under 65 \cite{demeesterNewEurotransplantKidney1998, Mayer2005} and the Eurotransplant Senior Program (ESP) for kidneys from donors aged 65 and older \cite{smitsEvaluationEurotransplantSenior2002}.
\par
Changes to ETKAS and ESP are regularly proposed, and are often based on medical insights or ethical considerations. An example of a medical insight is that the point system that forms the basis of ETKAS places equal emphasis on Human Leucocyte Antigen (HLA) matching on the A, B, and DR loci, which has been described to lead to additional mismatches on the DR locus \cite{demeesterNewEurotransplantKidney1998, vereerstraetenExperienceWujciakOpelzAllocation1998}. Based on findings that HLA-DR mismatches are most deleterious to graft survival \cite{Roberts2004, johnsonFactorsInfluencingOutcome2010}, proposals have been made to instead emphasize DR matching in ETKAS \cite{vereerstraetenAllocationCadaverKidneys1999, Doxiadis2004}. Other proposals are to make candidates under age 65 eligible for ESP \cite{Ssal2020}, to introduce HLA-DR matching in ESP \cite{deFijter2023}, to introduce candidate-donor age matching \cite{vonsamson-himmelstjernaContinuousDonorrecipientAge2024}, epitope matching \cite{Niemann2021}, and to give extra priority to candidates who are immunized, but do not meet entry for the Acceptable Mismatch (AM) program which is Eurotransplant's special program for highly immunized candidates \cite{ziemannUnacceptableHumanLeucocyte2017, zecherImpactSensitizationWaiting2022a, deferranteImmunizedPatientsFace2023}. 
\par
Within Eurotransplant, such areas of improvement are regularly discussed by the Eurotransplant Kidney Advisory Committee (ETKAC), which consists of an abdominal surgeon, an immunologist, and nephrologists who represent the Eurotransplant member countries. Despite ETKAC discussions on these topics, ETKAS and ESP have not changed much since their initiations in 1996 and 1999, respectively. An important barrier to updating the kidney allocation rules is that Eurotransplant has lacked tools which would enable Eurotransplant to quantitatively assess the effects of policy changes. Eurotransplant needs such tools to assess the adequacy of proposed rule changes, and to give national competent authorities insight into the negative consequences a policy change may have.
\par
Computer simulations of the kidney allocation system can fulfill this need, and are routinely used by other organ exchange organizations to introduce new allocation policies. For instance, in the United States, the Kidney-Pancreas Simulated Allocation Model (KPSAM) \cite{israniNewNationalAllocation2014} was used to revise allocation rules in 2014 \cite{israniNewNationalAllocation2014}, and this tool continues to be used for the proposal of further changes to allocation \cite{israniNewKidneyPancreas2021, mankowskiAcceleratingKidneyAllocation2019}. The allocation policies in France and the United Kingdom were also updated on the basis of bespoke computer simulations in 2015 and 2019, respectively \cite{jacquelinetChangingKidneyAllocation2006a, Audry2022, Mumford2018}. ETKAS itself was also originally designed based on computer simulations in 1993 \cite{wujciakComputerAnalysisCadaver1993,wujciakProposalImprovedCadaver1993a}.
\par 
This motivated us to develop a simulation toolbox which enables Eurotransplant and other stakeholders to quantify the impact of changes to ETKAS and ESP allocation rules. This tool, which we refer to as the ETKidney simulator, uses Discrete Event Simulation (DES) to mimic kidney allocation within Eurotransplant. The simulator was developed in close collaboration with medical doctors from Eurotransplant and presented to ETKAC, which has welcomed the ETKidney simulator as a tool to help inform policy discussions on kidney allocation. The Python code of the simulator is made publicly available together with synthetic data to enable collaborations with policymakers and scientists in evaluating alternative ETKAS and ESP allocation rules.\footnote{ \url{https://github.com/hansdeferrante/Eurotransplant_ETKidney_simulator}} In this paper, we describe how kidneys are allocated within Eurotransplant (Section \ref{sec:kidney_alloc}) and how this process is approximated by the simulator (Sections \ref{sec:design} and \ref{sec:modules}). Furthermore, we give insight into how closely the simulator approximates outcomes of ETKAS and ESP with input-output validation (Section \ref{sec:validation}). We also demonstrate how the ETKidney simulator can contribute to policy evaluation with three case studies (section \ref{sec:case_studies}). 

\section{Kidney allocation within Eurotransplant}
\label{sec:kidney_alloc}
\begin{figure}[h]
	\centering
	\resizebox{\linewidth}{!}{%
		\includegraphics{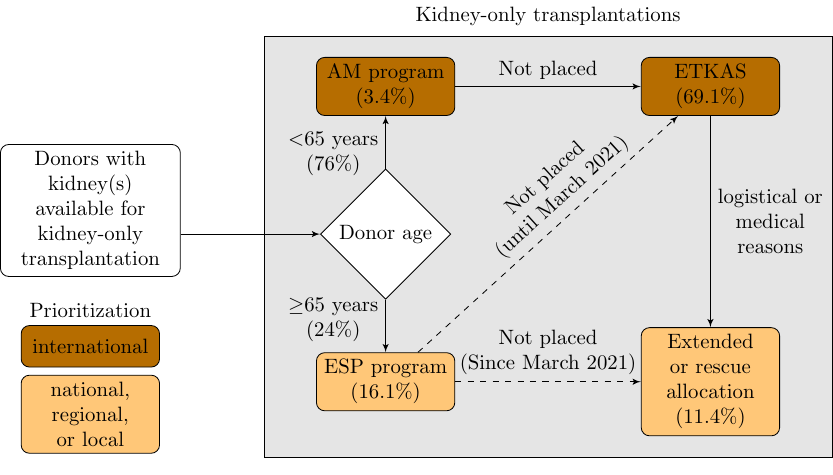}
	}
	\caption{Flow chart on how Eurotransplant offers kidneys for kidney-only transplantation. The percentages shown represent the proportion of kidney-only transplantations performed through each mechanism between 2014 and 2023 \cite{etStatsLibrary2072P}. Kidneys declined in ESP by all candidates were re-allocated via ETKAS until March 2021. Since March 2021, an extended allocation is triggered to place such kidneys.}
	\label{fig:flowchart_kidney_allocation}
\end{figure}

Figure \ref{fig:flowchart_kidney_allocation} shows how Eurotransplant allocates the deceased-donor kidneys which become available for kidney-only transplantation.\footnote{In case a multi-organ donor is reported to Eurotransplant, the donor's kidneys may first be accepted by candidates who wait for a combined transplantation of a kidney with another organ (heart, lung, pancreas, liver, or intestine). Such combined transplantations account for 7\% of all transplanted kidneys in Eurotransplant \cite{statlibrary2152P_AllET_Kidney}} Three standard allocation programs are used to place these kidneys: the Acceptable Mismatch (AM) program, ETKAS (\ref{subsection:etkas}), and ESP (\ref{subsection:esp}) (see Figure \ref{fig:flowchart_kidney_allocation}). Through which program Eurotransplant offers the kidney(s) is determined by the age of the donor. Kidneys from donors aged below 65 are first offered for transplantation through the AM program \cite{Heidt2015}, Eurotransplant's program for highly immunized candidates which accounted for 3\% of kidney-only transplantations between 2014 and 2023, and then through ETKAS, which accounted for 69\% of kidney-only transplantations. Kidneys from donors aged 65 and older are offered via ESP, which accounted for 16\% of kidney-only transplantations \cite{etStatsLibrary2072P}.
\par 
The remaining 11\% of kidney-only transplantations between 2014 and 2023 were the result of recipient-oriented extended allocation or competitive rescue allocation, which Eurotransplant refers to as \enquote{\textit{non-standard allocation}} mechanisms. Eurotransplant switches to these mechanisms under specific circumstances to prevent the loss of transplantable kidney(s) (see the Eurotransplant kidney manual \cite{manualKidney}). In either mechanism, centers located in the vicinity of the kidney(s) receive a \textit{center offer}, in which centers can propose candidates for transplantation. In extended allocation, centers have 60 minutes to propose candidates, and the proposed candidate(s) with the highest rank(s) on the original ETKAS or ESP match list receives the kidney(s) for transplantation. In rescue allocation, offers are competitive which means that the first center that responds receives the kidney.
\par
Not all kidneys offered by Eurotransplant are eventually transplanted. In fact, approximately 24\% of kidneys offered for transplantation between 2014 and 2023 were discarded \cite{etreport_1132P}.
\FloatBarrier
\subsection{Allocation based on match lists}
\label{subsection:matchlist_allocation}
Eurotransplant uses \textit{match lists} to determine which candidate receives a kidney offer at what moment. The composition and ordering of candidates on match lists are determined by program-specific \textit{eligibility}, \textit{filtering}, and \textit{ranking criteria}. A complete description of the program-specific criteria can be found in the Eurotransplant kidney manual \cite{manualKidney}, and a comprehensive summary is included in Appendix \ref{app:allocation rules}. 
\par 
Firstly, eligibility criteria determine whether a candidate is allowed to be included on the match list according to Eurotransplant rules. Examples of eligibility criteria are that the candidate must have the same blood group as the donor, and that the donor's HLA typing cannot include an HLA antigen which the center has reported as unacceptable for the candidate. Such \enquote{\textit{unacceptable antigens}} are used by centers to indicate that their candidate has a pre-existing immunization against specific HLA antigens, for instance because of a prior blood transfusion, pregnancy, or transplantation. Centers want to avoid offers with such antigens, because transplantation in the presence of a pre-existing immunization is likely to result in an early loss of the transplanted kidney.
\par 
The filtering criteria are specified by the transplantation centers, and determine whether Eurotransplant contacts the center to make an offer for the transplantation of a specific candidate. The filtering criteria can be distinguished into:
\begin{enumerate}[noitemsep]
	\itemsep0em
	\item \textit{allocation profiles}, with which centers can indicate that their patient does not want to receive offers from donors with certain characteristics (for instance, donors above a certain age, virology, etc.), and
	\item \textit{HLA mismatch criteria}, with which centers can specify minimal requirements for the HLA match quality between the donor and candidate on the HLA-A, -B and -DR loci.
\end{enumerate}
\par
Finally, ranking criteria determine the position of a candidate on the match list. This match list order determines the sequence in which candidates receive an offer. To determine this order, candidates are first grouped in match \textit{tiers} based on medical and ethical criteria (for instance, 0 HLA mismatches, pediatric status, and the location of the candidate relative to the donor). Candidates ranked in higher tiers receive absolute priority over candidates in lower tiers. Within tiers, candidates are ranked by points. The specific ranking criteria for ETKAS and ESP are described in Section \ref{section:match_list_and_matchpoints}.
\par 
In allocating kidneys, Eurotransplant starts by contacting centers with offers for specific, named candidates. Such offers are only made to candidates who appear on the \textit{filtered match list}, which means that candidates must meet the program's eligibility and filtering criteria. 
In extended allocation, centers are additionally allowed to select candidates from \textit{unfiltered match list}, which means that candidates only have to meet the program's eligibility criteria. In competitive rescue allocation, centers are also allowed to propose blood group compatible candidates who did not appear on the match list.


\subsection{The ETKAS program}
\label{subsection:etkas}
In ETKAS, kidneys from donors aged below 65 are offered to candidates of any age. The point system of ETKAS was designed based on computer simulations in 1993 \cite{wujciakComputerAnalysisCadaver1993, wujciakProposalImprovedCadaver1993a}, and penalizes HLA mismatches between the donor and the candidate because of its negative impact on kidney graft survival \cite{demeesterNewEurotransplantKidney1998}. To avoid extensive waiting times for candidates for whom a high-quality HLA match is rare, the \WOP\ also awards mismatch probability points and points for the time a candidate has waited on dialysis. Distance points are awarded to promote local and domestic transplantations, and balance points are awarded to maintain a balance in the international exchange of kidneys \cite{demeesterNewEurotransplantKidney1998}. Finally, the point system awards points to sensitive patient groups: pediatric candidates receive HLA bonus points and pediatric points, and candidates with the High Urgency (HU) status receive HU points. 

\subsection{The Eurotransplant Senior Program (ESP)}
\label{subsection:esp}
ESP was initiated as an old-for-old program in 1999, is used to offer kidneys from donors aged 65 and older with priority to candidates aged 65 and older. The aims behind ESP were (i) to increase the number of donors aged 65 and older, (ii) to improve quality of life for older candidates, and (iii) to decrease the cold ischemia times for older donor kidneys \cite{eurotransplantinternationalfoundationMinutesETKACMeeting1998}. These aims were achieved by only offering kidneys to candidates located in the vicinity of the donor, and prioritizing candidates solely based on accrued dialysis time \cite{smitsEvaluationEurotransplantSenior2002,Frei2008}. 
\par
Since the program's introduction in 1999, ESP has changed in several ways. One change is that prospective HLA typing of the donor was not performed in ESP in order to maximally reduce cold ischemia times \cite{smitsEvaluationEurotransplantSenior2002}. Because this meant that the donor's HLA typing was unknown at time of allocation, candidates were transplanted without regard to HLA match quality and immunized candidates were not eligible to receive offers via ESP. Since 21-01-2025, all Eurotransplant countries require the donor's HLA typing to be available before kidneys can be allocated via ESP. This has made it possible to also offer kidneys to immunized candidates via ESP. Another important change is that kidneys not accepted on the filtered match list in ESP were re-allocated via ETKAS. This resulted in extended allocation times, because most such offers were declined because of geographical distances and/or age mismatch. Since March 2021, unplaced ESP kidneys are therefore instead allocated via non-standard allocation based on the \textit{unfiltered} ESP match list. To ensure that non-local candidates and candidates aged below 65 still have access ESP kidneys, such candidates can appear on the unfiltered ESP match list since March 2021 (see Appendix \ref{app:allocation rules}).

\section{Purpose and design of the ETKidney Simulator}
\label{sec:design}
The purpose of the ETKidney simulator is to enable Eurotransplant to assess the impact of changes to ETKAS and ESP allocation rules on waiting list outcomes. To meet this goal, we use Discrete Event Simulation (DES) to simulate kidney allocation according to the actual ETKAS and ESP allocation rules implemented in March 2021. Simulation of the AM program is beyond the scope of the ETKidney simulator, because definition of acceptable antigens is based on an individualized risk-benefit analysis which requires specialized immunological knowledge. The simulator was also not designed for the analysis of kidney discard rates.
\par
Relevant system states are (i) the statuses of candidates who wait in ETKAS or ESP for a kidney-only transplantation, and (ii) the export balances of Eurotransplant member countries, which affect a candidate's rank on ETKAS match lists because of Eurotransplant's balance point system (see Section \ref{subsection:etkas}). In DES, we study how such system states evolve in response to a series of discrete events. For the ETKidney simulator, we distinguish between three types of events:
\begin{enumerate}[noitemsep]
	\item updates to candidate states, which are changes to a candidate's waiting list status (transplantable, non-transplantable, HU, removed, transplanted, or waiting list death), their allocation profile, their unacceptable antigens, their HLA mismatch criteria, the reporting of an antibody screening, or a choice for ETKAS or ESP in Germany (see Appendix \ref{app:allocation rules}),
	\item the arrivals of donors with one or two kidneys available for kidney-only transplantation through ETKAS or ESP, which generally results in a transplantation, and 
	\item transplantations across country borders through allocation programs other than ETKAS and ESP (AM and combined transplantations), which count towards an Eurotransplant member country's export balance.
\end{enumerate}
\par
\noindent

\subsection{Input data required for the ETKidney simulator}
\label{subsection:input_streams}
Users of the ETKidney simulator have to specify the input streams which define the candidate and donor events. Furthermore, an input stream of international transplantations can be specified, which is used to initialize ETKAS balances and to define balance update events. These input streams can be based on historic data, synthetic data, or combinations thereof.
\par
For candidates and donors, the data in the input streams must include all administrative and medical information required by the eligibility, filtering, and ranking criteria. 
Additional information may be required by the graft offer acceptance module to simulate kidney offer acceptance behavior, and by the post-transplant module to simulate post-transplant survival.
\par 
For candidates, the input streams must include complete information on what would happen to each candidate until they exit the waiting list because of a waiting list removal or a waiting list death. An implication of this requirement is that candidate input streams cannot be solely based on historic registry data: after all, waiting list removals or waiting list deaths are not observed for candidates who were transplanted via ETKAS or ESP.
\par
For simulations done in this paper, we use input streams based on historic data. A counterfactual status imputation procedure is used to impute what would have happened on the waiting list to candidates who were transplanted through ETKAS or ESP. This imputation procedure was modified from a procedure originally devised for development of the ELAS simulator, a simulator developed for Eurotransplant liver allocation \cite{deferranteELASSimulator2023}. In summary, the procedure consists of:
\begin{enumerate}[noitemsep]
	\item constructing a risk set of candidates who are similar to the transplanted candidate in background characteristics and expected remaining survival time, but who are still waiting for a transplantation,
	\item randomly sampling a candidate from this risk set, and copying over their future status updates to the transplanted candidate.
\end{enumerate}
By repeating this procedure, several potential status update trajectories can be constructed for a candidate. We describe this counterfactual status imputation procedure in detail in Appendix \ref{app:imp_proc}.

\subsection{Initialization of the ETKidney simulator's system state}
\label{subsection:initialization}
A simulation settings file is used to specify a simulation start and end date (which jointly define the simulation window), paths to the input streams, and paths where outcomes of the simulation (transplantation, patient states) are written to when the simulation terminates. The system states are initialized based on these simulation settings.
\par
The balance system is initialized by loading and processing all international transplantations which occurred before the simulation start date from this input stream. Additionally, the simulator schedules a balance update event for cross-border transplantations which occur within the simulation window via combined transplantations or via the AM program. This ensures that ETKAS may be simulated under realistic kidney balances.
\par 
From candidate input streams, the simulator loads all candidates who have an active waiting list status within the simulation window. Listings of repeat transplantation candidates whose initial transplantation falls within the simulation window are excluded, which is necessary to ensure that a candidate cannot simultaneously wait for a primary and repeat transplantation in ETKidney simulation runs. 
Status updates of candidates which occurred before the simulation start date are processed as part of the initialization procedure, which ensures that the state of candidates coincides with the candidate's actual state on the simulation start date. Finally, for each candidate a single patient event is scheduled in the Future Event Set (FES) timed at the first available status update of the candidate.
\par 
From donor input streams, all donors reported during the simulation window are loaded. For each donor, a donor event is scheduled in the FES on the date the donor was reported to Eurotransplant.

\subsection{Overview of the simulation}
\label{subsection:overview}
\begin{figure}[h]
	\centering
	
	\resizebox{1\linewidth}{!}{%
		\includegraphics{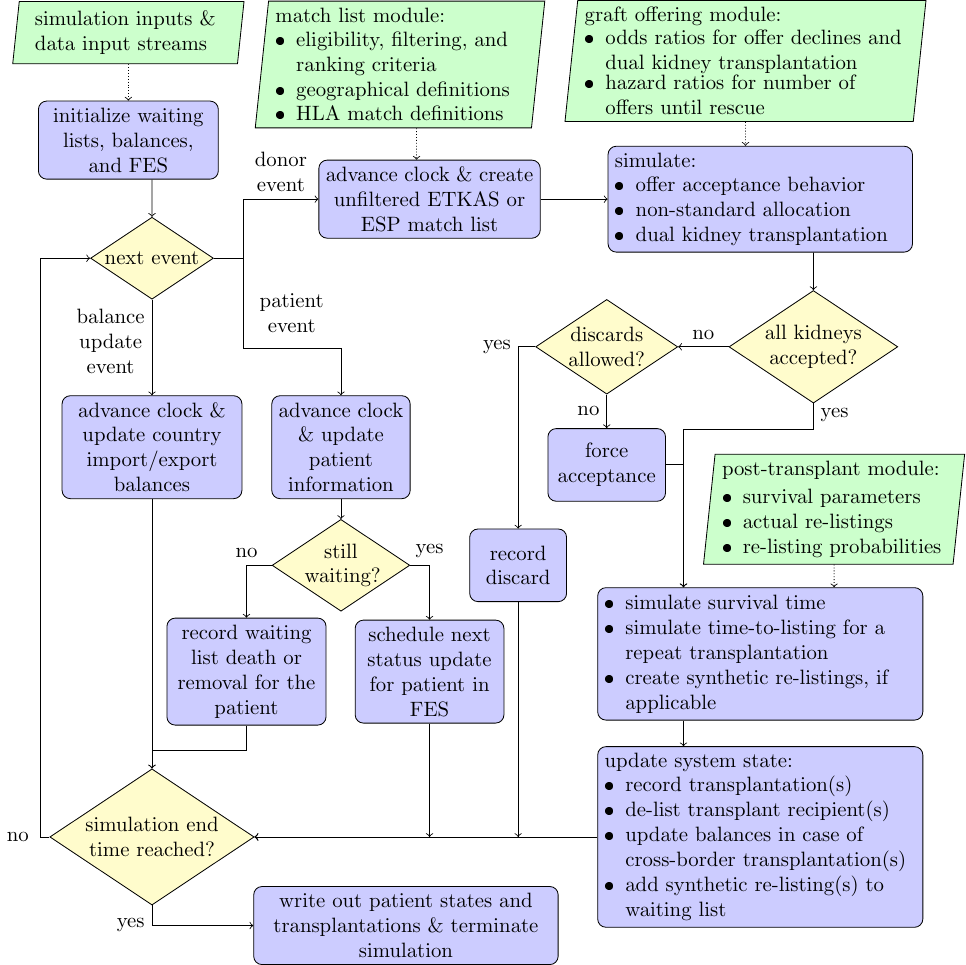}
	}
	\caption{Event handling flowchart for the ETKidney simulator. Inputs and parameters are represented using parallelograms. In case a kidney was declined by all candidates, simulation settings determine whether a discard is recorded or whether a candidate is forced to accept the kidney for transplantation. FES: Future Event Set.}
	\label{fig:eventhandling_flowchart}
\end{figure}

Figure \ref{fig:eventhandling_flowchart} illustrates how events are processed in the simulation. Balance events encode an international transplantation external to ETKAS or ESP, and are handled by simply updating the import/export balances of involved countries. Patient events encode a change to the state of a patient, and the simulator handles such events by updating the state of the corresponding candidate. Handling of donor events is more elaborate, because allocation of the kidney(s) through ETKAS or ESP has to be approximated by the simulator.

\par
To approximate such kidney-only allocation, the ETKidney simulator's \textit{match list module} was developed (see Section \ref{section:match_list_and_matchpoints}). This module first creates, depending on the age of the donor, an unfiltered ETKAS or ESP match list. This match list contains for each candidate who meets the corresponding program's eligibility criteria a \textit{match record}. The ESP / ETKAS match list is automatically ordered based on the respective program's ranking criteria, with the number of points awarded to each match record determined by the \textit{point score module} (see Section \ref{section:match_list_and_matchpoints}). 
\par
Based on the ordered match list, the \textit{graft offering} module simulates which candidate accepts the kidney(s) (see Section \ref{section:acceptance}). In rare cases, all candidates who appear on the match list are simulated to decline the kidney offer. In case this happens for an ESP match list, the graft offering module will try to place the kidneys through non-standard ESP allocation (which is how Eurotransplant places kidneys from donors aged 65 and older since March 2021, see Section \ref{subsection:esp}). In the very rare case that kidney(s) remain unplaced, the graft offering module can either (i) record a discard for the kidney(s), or (ii) force transplantation of the candidate who had the highest predicted probability of accepting the offer. For the simulations in this paper, option (ii) is used because the donor input stream used for simulation consists only of donors whose kidneys were actually transplanted through ETKAS and ESP.
\par
Transplantations are then recorded for candidates who were simulated to accept the kidneys, and the kidney balances are updated in case of a cross-border transplantation. The \textit{post-transplant module} simulates post-transplant outcomes for transplant recipients, and schedules for candidates who are simulated to list for repeat transplantation before the simulation termination date a \textit{synthetic re-listing} (see Section \ref{section:synthetic_re-registration}).
\FloatBarrier
\section{Modules of the ETKidney simulator}
\label{sec:modules}
The section describes key modules of the ETKidney simulator: the HLA system module (Section \ref{section:hla_system}), the balance system module (Section \ref{section:balance_system}), the match list and point system module (Section \ref{section:match_list_and_matchpoints}), the graft offering module (Section \ref{section:acceptance}), and the post-transplant module (Section \ref{section:posttransplant}).
\label{section:modules}

\subsection{The HLA system module}
\label{section:hla_system}
The HLA system module implements all mechanisms through which Eurotransplant prioritizes HLA matching. These procedures are (i) determining how many mismatches there are at HLA loci of interest, (ii) calculation of virtual Panel-Reactive Antibodies (vPRA) on the basis of unacceptable antigens, and (iii) calculation of mismatch probability points. 
\subsubsection{Calculation of HLA mismatches}
\label{subsection:hla_mm}
The HLA system module can determine for a given patient HLA and given donor HLA how many antigen mismatches there are per locus (0, 1, or 2). At which loci HLA mismatches are to be determined is stored in the simulation settings file. By default, the module only counts mismatches for the HLA-A, -B, and -DR loci, because the current \WOP\ only awards points for such mismatches. HLA-A and HLA-B mismatches are determined at the level of broad antigens, HLA-DR mismatches at the level of split antigens (see \cite{manualKidney}).

\subsubsection{Virtual Panel-Reactive Antibodies (vPRA)}
\label{subsection:vpra}
Transplantation centers can report HLA antigens as unacceptable for their patients. With unacceptable antigens, patients have fewer transplantation opportunities because they will be ineligible for kidney offers from donors with unacceptable antigens. The relative restriction in access is quantified by Eurotransplant with virtual Panel-Reactive Antibodies (vPRA), which is defined as the percentage of the donor pool whose HLA typing includes unacceptable antigens. The vPRA affects a candidate's position on the match list, because the vPRA is used to calculate a candidate's mismatch probability, for which ETKAS awards points. For allocation, the vPRA is quantified against the ETRL donor panel, which is a database of 10,000 donors which were recently reported in the Eurotransplant area, which is maintained by the Eurotransplant Reference Laboratory (ETRL). The HLA system module can also quantify the vPRA against a user-specified input database of 10,000 donor HLAs.

\subsubsection{Calculation of the Eurotransplant mismatch probability (MMP)}
\label{subsection:mmp}
The mismatch probability quantifies for each candidate the chance that among the next 1,000 donors reported to Eurotransplant there is at least one donor who is favorably matched with the candidate. Such a favorably matched donor is defined as a donor who (i) is blood group identical with the candidate, (ii) has no HLAs unacceptable for the candidate, and (iii) has at most 1 HLA-ABDR mismatch with the candidate \cite{wujciakMatchabilityImportantFactor1997}. Eurotransplant calculates this mismatch probability analytically as:
$$\texttt{MMP} = \Big[1 - f_{BG}\cdot (1-\texttt{vPRA})\cdot p_{\leq1mm}\Big]^{1000},$$
where $f_{BG}$ is the candidate's blood group frequency and $p_{\leq1mm}$ is a quantification of the probability that a donor has at most 1 HLA-ABDR mismatch with the candidate. Candidates with a higher mismatch probability thus have relatively fewer favorably matched donors. To compensate candidates for this, the \WOP\ awards points based on this mismatch probability.
\par
We point out that $p_{\leq1mm}$ is calculated analytically based on HLA frequencies \cite{manualKidney}, which ignores haplotype linkage disequilibrium \cite{wujciakMatchabilityImportantFactor1997}. This was implemented as the default option for the HLA system module, but the module can also quantify $p_{\leq1mm}$ by counting the fraction of donors with at most 1 HLA-ABDR mismatch among the same donor pool of 10,000 donors used to calculate vPRAs. We refer to this latter quantity as the \enquote{\textit{1-ABDR HLA mismatch frequency}} (and use this quantity in the second case study, see section \ref{sec:application_vpra}).

\subsection{The balance system module}
\label{section:balance_system}
Eurotransplant's member countries have varying organ donation rates. To ensure that countries with higher organ donation rates also have increased access to transplants, Eurotransplant strives towards a balance in the international exchange of organs. For kidneys, Eurotransplant keeps track of the \textit{net kidney export balances} of the member countries, and awards points in ETKAS based on these balances.\footnote{The number of balance points awarded is calculated by subtracting from each member country's export balance the export balance of the country which is the largest importer (a negative number), and multiplying the outcome by the balance weight} Within Austria, an additional regional balance system exists to ensure that the Austrian centers benefit equally from cross-border transplantations \cite{manualKidney}.
\par
The balance system module implements both these national and regional balance systems for the ETKidney simulator. Initialization of the net exports is possible, and can be based on historic data. By default, separate balances are maintained for each donor age group (0-17 years, 18-49 years, 50-64 years, or 65+), which coincides with the balance system introduced on 01-04-2019 \cite{manualKidney}. 

\subsection{The match list module \& point system module}
\label{section:match_list_and_matchpoints}
The match list module is used to create ETKAS or ESP match lists, which serve as input for the graft offering module. Only candidates who meet the corresponding program's eligibility criteria appear on match lists, with the rank of a candidate based on \textit{tiers} and \textit{points}. Candidates ranked in higher tiers receive absolute priority over candidates in lower tiers. Within tiers, candidates are ranked by points. In ETKAS, tiers are in order of descending priority:
\begin{enumerate}[noitemsep]
	\item candidates with 0 HLA-ABDR mismatches with the donor,
	\item offers of pediatric kidneys to pediatric candidates,
	\item all other candidates.
\end{enumerate}
The point system for ETKAS awards:
\begin{enumerate}[noitemsep]
	\item 33.33 points per year of accrued dialysis time,
	\item 400 minus 66.66 points per HLA-ABDR mismatch. These points are doubled if the candidate is pediatric,
	\item 100 points if the candidate is pediatric,
	\item 500 points if the candidate has the High Urgency (HU) status, 
	\item up to 100 mismatch probability points,
	\item balance points, the amount of which is based on net export balances,
	\item up to 300 distance points if the candidate is located in the same country as the donor (the specific amount depends on the country, see \cite{manualKidney}).
\end{enumerate}
\par
ESP has used nine tiers since March 2021, which are defined based on the location of the candidate relative to the donor and whether the candidate is aged 65 and older. Within ESP tiers, points are awarded based on accrued dialysis time (see Appendix \ref{app:allocation rules} and \cite{manualKidney}).

\subsection{The graft offering module}
\label{section:acceptance}
\begin{figure}[h]
	\centering
	\resizebox{.9\linewidth}{!}{%
		\includegraphics{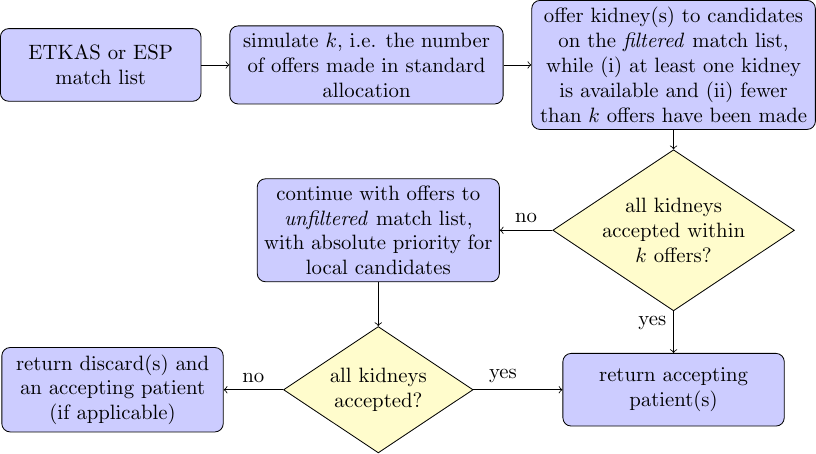}
	}
	\caption{Summary of how Eurotransplant's kidney offering process is approximated by the ETKidney simulator.}
	\label{fig:flowchart_graft_offering}
\end{figure}

The graft offering module mimics how Eurotransplant offers kidneys to centers for transplantation, and returns based on a match list which candidates accept the kidney for transplantation (if any). How kidney offers are simulated in the ETKidney simulator is illustrated in Figure \ref{fig:flowchart_graft_offering}. To mimic Eurotransplant's allocation process, the module first simulates based on the donor how many offers are maximally made in standard allocation (see Section \ref{subsubsection:extended_or_rescue}). We denote this maximum number of offers by $k$. The module then makes kidney offers to candidates in order of their ranking on the \textit{filtered} match list, either until $k$ offers have been made or until all available kidneys have been accepted for transplantation. How center offer acceptance behavior is simulated is described in Section \ref{subsubsection:two_stage_acceptance}. 
\par 
\par
If not all kidneys have been accepted after $k$ offers in standard allocation, the module re-orders match lists with absolute priority for candidates who are located in the vicinity of the donor. This procedure was implemented to approximate the outcomes of non-standard allocation. At this point, the module also makes offers to candidates who appear only on the \textit{unfiltered} match list, because centers can select such candidates in extended and rescue allocation (see Section \ref{subsection:matchlist_allocation}). Offers in non-standard allocation are made until all kidneys have been accepted, or until the match list is exhausted. The module then returns the patients who have accepted the kidneys, as well as how many kidneys were not successfully placed.
\par
In particular circumstances, centers are allowed to accept both kidneys for transplantation. The graft offer acceptance module also simulates whether such dual kidney transplantations are performed when a kidney is accepted (see Section \ref{subsubsection:dkt}).

\subsubsection{Triggering of non-standard allocation}
\label{subsubsection:extended_or_rescue}
The graft offering module switches to non-standard allocation after $k$ declines. We modelled $K$ with a Cox proportional hazards model with adjustment for donor characteristics (donor age, hepatitis, death cause, last creatinine, diabetes, smoking, proteinuria, blood group, and extended donor criteria). Cox models were stratified by the program (ETKAS / ESP), and within ETKAS additionally by the donor country. 
Default parameters and baseline hazards to simulate from this model are made available with the ETKidney simulator. These parameters were estimated on match lists from 01-01-2014 to 01-01-2024. ESP match lists from before March 2021 were excluded for estimation of these parameters, because 2021 changes to the ESP program (see Section \ref{subsection:esp}) have substantially increased the frequency of non-standard allocation in ESP.

\subsubsection{A two-staged kidney offer acceptance procedure}
\label{subsubsection:two_stage_acceptance}
Transplantation centers frequently decline offers received from Eurotransplant, for instance because of concerns about donor or kidney quality, logistical reasons, or recipient reasons. When centers receive a kidney offer from Eurotransplant in standard allocation for a specific candidate, the center can decide to (i) accept the kidney for the transplantation of the named candidate, (ii) decline the kidney for the named candidate while remaining open for offers to other candidates, or (iii) decline the kidney for the entire center. 
\par 
To approximate this decision-making process in the graft offering module, a two-staged offer acceptance procedure was implemented. This procedure:
\begin{enumerate}[noitemsep]
	\itemsep0em
	\item simulates \textit{center-level} decisions based on donor characteristics (donor death cause, age, last creatinine, blood group, heart-beating or non-heart-beating donation, and extended donor criteria) and center characteristics (country, distance to the donor). These models simulate whether a center is willing to accept the kidney(s) for transplantation for any candidate in the center, and
	\item simulates \textit{patient-level} decisions to determine whether the center accepts a kidney offer for a specific candidate. These decisions are simulated only if the center is willing to accept the donor, and are simulated based on donor characteristics (see above), candidate characteristics (age, pediatric status, HU status, vPRA, dialysis time, prior kidney transplantation), and match characteristics (HLA match, geographic distance, candidate-donor age difference).
\end{enumerate}
Logistic models are used to simulate both these decisions. Because HLA match quality has historically been ignored in ESP and because ESP and ETKAS have different donor and patient populations, we anticipated that kidney offer acceptance behavior would differ between ETKAS and ESP. Therefore, separate models are used to simulate offer acceptance decisions within ETKAS and ESP.
\par
The odds ratios used for simulations are made publicly available with the ETKidney simulator, and were estimated on historic ETKAS / ESP match lists with mixed effect logistic regressions. For ETKAS, odds ratios were estimated on historical data between 01-01-2012 and 01-01-2021. For ESP, odds ratios were estimated between 01-01-2018 to 01-01-2024. We included data from after 2021 to estimate the ESP odds ratios because ESP offers can only be made to candidates younger than 65 and across country borders since March 2021 (see Section \ref{subsection:esp}). Odds ratios were estimated with random effects to account for within-donor, within-candidate and within-center correlations in organ offer acceptance decisions. 

\subsubsection{Simulation of dual kidney transplantations}
\label{subsubsection:dkt}
Annually, approximately 20 to 30 candidates receive a dual kidney transplantation, which is allowed in specific circumstances (for instance, because of anatomical reasons, see \cite{manualKidney}). The graft offering module simulates whether such a dual kidney transplantation is performed based on donor and patient characteristics (candidate age, donor age, country of listing, and rescue allocation). For this, a logistic model is used. The odds ratios made available with the ETKidney simulator were estimated on kidney offers made between 01-01-2018 and 01-01-2024 where both kidneys were still available for transplantation.

\subsection{The post-transplant module}
\label{section:posttransplant}
In total 13\% of candidates registered on the kidney transplantation waiting list have had a prior kidney transplantation, and nearly 3\% of kidney transplant recipients enlist for a repeat kidney transplantation within 1 year of transplantation \cite{eurotransplantinternationalfoundationWaitingListRegistrations2024}. To simulate such post-transplant survival and listing for repeat transplantation, the post-transplant module was implemented. This module simulates a post-transplant failure time $t$ and potential time-to-relisting $r$ relative to $t$ for each transplant recipient (section \ref{section:failure_and_re-registration_time}). If the candidate is simulated to enlist for a repeat transplantation before the simulation end date, the module creates a synthetic re-listing for transplant recipients who enlist for a repeat transplantation within the simulation window (see section \ref{section:synthetic_re-registration}).

\subsubsection{Simulation of post-transplant failure time and a potential re-listing date}
\label{section:failure_and_re-registration_time}
For each kidney transplantation, the post-transplant module simulates a patient failure time $t$ based on donor characteristics (age, non-heart-beating donation, last creatinine, death cause, extended donor criteria), patient characteristics (age, dialysis time, repeat kidney transplantation, country of listing), and match characteristics (HLA match, acceptance reason, year of transplantation, international transplantation). This failure time is defined as a post-transplant death or a repeat kidney transplantation, whichever occurred first. We modelled $T$ with a Weibull model based on recipient, donor, and transplantation characteristics. Details on this model is included in Appendix \ref{app:posttxp}. The scale and shape parameters supplied with the ETKidney simulator were estimated based on ETKAS and ESP transplantations from 01-01-2011 and 01-01-2021.
\par
Most transplant recipients enlist for a repeat kidney transplantation before a patient failure would materialize. The ETKidney simulator approximates this by also simulating a time-to-relisting $r$. Because a potential re-registration logically has to occur before a patient death or re-transplantation, we simulate the time-to-relisting $r$ based on the empirical distribution of $R$ relative to $T$. We stratified these distributions by candidate age groups and time-until-failure $t$ (details are included in  Appendix \ref{app:posttxp}). 
\subsubsection{Creating synthetic re-listings}
\label{section:synthetic_re-registration}
In case transplant recipients are simulated to enlist for a repeat transplantation before the simulation end date, the post-transplant module creates a \textit{synthetic re-listing} for this candidate. This synthetic re-listing is created by combining the static information from the transplant recipient with the dynamic candidate status updates from an actually re-listed candidate. The actually re-listed candidate is chosen such that they are similar to transplant recipient in terms of background characteristics as well as in time-to-failure $t$. Details on this procedure are included in Appendix \ref{app:posttxp}.
\par
A challenge for creating a synthetic re-listing is that kidney transplantation is a strongly immunizing event \cite{Lopes2015}. Many repeat transplantation candidates will therefore have developed de novo Donor-Specific Antibodies (dnDSAs), which centers can report as unacceptable. Candidates who are listed for repeat transplantation therefore tend to have increased vPRAs and reduced access to kidneys. It is not clear how to simulate such de novo immunization, because (i) HLAs are cross-reactive, which means that centers frequently also report HLAs as unacceptable which were not present in the donor \cite{Lucas2015}, (ii) the immunogenicity of donor HLAs also depends on the patient's own HLA \cite{Lucas2015}, and (iii) HLA laboratories and transplantation centers have different attitudes in labeling HLAs of the initial donor as unacceptable \cite{Ssal2013, Ziemann2022}. 
\par 
Accurate simulation of de novo immunization is thus beyond the scope of the ETKidney simulator. Instead, a very simple procedure was implemented, which is to assume that candidates have a fixed probability of becoming immunized against any mismatched donor antigen. By default, this probability is set to 20\%. This probability was chosen because having one additional mismatch per locus increases the probability of reporting unacceptable antigens against that specific locus on average with 10 to 25\% (depending on the locus) \cite{Isaacson2022}.

\section{Verification and validation}
\label{sec:validation}
This section describes verification and validation efforts taken to ensure that the ETKidney simulator closely mimics ETKAS and ESP. Under model verification, we understand efforts taken to \textit{\enquote{ensure that the computer program of the computerized model and its implementation are correct}} \cite{sargent2020}. Under model validation, we understand efforts taken to assess that the model \textit{\enquote{possesses a satisfactory range of accuracy consistent with the intended application of the model}} \cite{sargent2020}. 

\subsection{Verification of the ETKidney simulator}
\label{section:verification}
We built unit tests to ensure that the behavior of the ETKidney's simulator modules is in agreement with their intended behavior. For instance, unit tests were constructed to check whether HLA match qualities returned by the HLA system module matched HLA match qualities of actual ETKAS match lists. Unit tests were also used to ascertain that the HLA system module returned the correct mismatch probabilities and vPRAs. 

\par
Simulation of the graft offering process and of post-transplant survival is based on statistical models, which were estimated in R. Unit tests were constructed to ensure that predicted probabilities in the ETKidney simulator matched predicted probabilities in R for offer acceptance decisions and for post-transplant survival.

\subsection{Validation of the ETKidney simulator}
\label{section:validation}
To ensure face validity of the model, medical doctors from Eurotransplant were actively involved in the development and conceptual design of the simulator. We also had meetings with ETKAC and the ETRL, and we presented the model at the Eurotransplant Annual Meeting to collect feedback on the conceptual model from other major stakeholders, such as medical doctors and transplantation coordinators from the transplantation centers.
\par 
We use input-output validation to assess how closely the model can approximate outcomes of ETKAS and ESP. For this, we simulate kidney allocation between 01-04-2021 and 01-01-2024 under the actual allocation rules used within this simulation window. For the donor input stream, we used all 4,326 donors with kidneys transplanted through ETKAS or ESP in the simulation period. For the candidate input stream, we used all 9,589 candidates who were on the waiting list on 01-04-2021, and all 14,333 candidates who were activated on the waiting list in the simulation period. To enable accurate simulation of the ETKAS balances, international transplantations in the AM program or via combined transplantations were exported from the Eurotransplant database. In simulations, we schedule donors, candidate status updates, and balance update events on the dates these were actually reported to Eurotransplant. Our input-output validation exercise thus keeps the inputs as close as possible to reality, and assesses whether outputs of the ETKidney simulator are comparable to the actual outputs of ETKAS and ESP.
\par
Important is that outputs of the simulator depend on several stochastic processes (offer acceptance behavior, re-listing, and non-standard allocation). To give insight into the resulting variability in simulator outputs, we simulate ETKAS and ESP allocation 200 times over the simulation window and report \enquote{95\%-interquantile ranges} for relevant summary statistics. These 95\%-IQRs are obtained by simulating allocation 200 times and reporting the 2.5th and 97.5th percentiles of simulation outputs. For each of these 200 simulation runs, we use a different set of imputed status trajectories (see Section \ref{subsection:input_streams}). We say that the ETKidney simulator is \textit{well-calibrated} for a quantity of interest if the actually observed summary statistic falls within the 95\%-IQR of the 200 simulations. We do not test statistically whether the mean outcome over the simulations is different from the actually observed outcome, because such tests can always be made statistically significant by increasing the number of simulation runs.

\subsubsection*{Results of input-output validation}
\label{subsection:val_wl_outc}
Table \ref{tab:waitlist_validation} reports input-output validation results for outcomes on the kidney waiting list. The ETKidney simulator is well-calibrated for almost all summary statistics: the total number of transplantations, the number of dual kidney transplantations, the number of ETKAS / ESP transplantations, the number of re-listings, and the number of waiting list deaths per country in all countries. We only observe miscalibration for the number of waiting list deaths in Hungary (-11\%) and the active waiting list size at simulation termination (+1.8\% too many candidates have an active waitlist status).
\begin{table}[h]
	\caption{Input-output validation of waiting list outcomes between 01-04-2021 to 01-01-2024. For simulations, the numbers shown are averages and 95\%-IQR of outcomes over 200 simulations. Ranges are displayed in bold if the simulator is not well-calibrated, i.e. if the actual statistic does not fall within the 95\%-IQR.}
	\hspace*{-.05\linewidth}
\centering
\begin{tabular}{L{4.9cm} R{5.2cm} L{2cm}}
\toprule
\multirow[b]{1.5}{*}{category} & \multirow[b]{1.5}{*}{\makecell{simulated results\\(average and 95\%-IQR)}} & \multirow[b]{1.5}{*}{\makecell{actual data\\(2021-2024)}}\\
& & \vspace{-0em} \\
\midrule
\addlinespace[0.3em]
\multicolumn{3}{l}{\textbf{transplantations through ETKAS or ESP}}\\
\hspace{1em}number of unique donors & 4326 & 4326\\
\hspace{1em}number of transplantations & 7546   [7530-7560] & 7549\\
\addlinespace[0.3em]
\multicolumn{3}{l}{\textbf{number of transplantations by type}}\\
\hspace{1em}single kidney & 7471   [7440-7500] & 7484\\
\hspace{1em}dual kidney & 74     [60-90] & 65\\
\addlinespace[0.3em]
\multicolumn{3}{l}{\textbf{number of transplantations by allocation mechanism}}\\
\hspace{1em}ESP & 1740   [1727-1752] & 1745\\
\hspace{1em}ETKAS & 5805   [5797-5814] & 5804\\
\addlinespace[0.3em]
\multicolumn{3}{l}{\textbf{waiting list}}\\
\hspace{1em}initial active waiting list & 9589 & 9589\\
\hspace{1em}number of listings & 24028  [24010-24046] & 24038\\
\hspace{1em}re-listings in simulated period & 106    [88-124] & 116\\
\hspace{1em}final active waiting list size & \textbf{10142  [\Plus2\%, 10095-10194]} & 9958\\
\hspace{1em}waiting list removals (count) & 1449   [1427-1473] & 1469\\
\hspace{1em}waiting list deaths (count) & 1146   [1122-1165] & 1140\\
\addlinespace[0.3em]
\multicolumn{3}{l}{\textbf{number of waiting list deaths by country}}\\
\hspace{1em}Austria & 73     [67-80] & 73\\
\hspace{1em}Belgium & 82     [73-90] & 73\\
\hspace{1em}Croatia & 40     [36-46] & 40\\
\hspace{1em}Germany & 723    [705-740] & 713\\
\hspace{1em}Hungary & \textbf{128    [-11\%, 121-135]} & 144\\
\hspace{1em}Netherlands & 92     [84-99] & 90\\
\hspace{1em}Slovenia & 7      [5-10] & 7\\
\bottomrule
\end{tabular}

	\label{tab:waitlist_validation}
\end{table}
\par
The left side of Table \ref{tab:transplant_validation} reports input-output validation results for ETKAS transplantations. The simulator is well-calibrated for the number of transplantations placed by allocation mechanism (standard or non-standard), transplantations by candidate age group, and transplantations in repeat transplantation candidates. The simulator is also well-calibrated for the number of transplantations by HLA match quality, with only the number of 0-mismatched transplantations overestimated ($\Plus$5\%). The number of transplantations in candidates with vPRAs exceeding 95\% is underestimated (-17\%). Such miscalibration seems to have been the result of the introduction of the virtual crossmatch in January 2023 (see Supplementary Table \ref{tab:vpra_vxm}), potentially because the number of positive recipient center crossmatches has decreased after introduction of the virtual crossmatch \cite{Heidt2024}. The simulator is also well-calibrated for the number of transplantations per country, with only a slight overestimation observed in Croatia ($\Plus 3\%)$ and a slight underestimation observed in Hungary (-2\%). Geographical sharing within ETKAS is underestimated, with 5\% extra local or regional transplantations, 11\% fewer interregional and 11\% fewer international transplantations.
\par
The right side of Table \ref{tab:transplant_validation} shows validation results for ESP transplantations. The simulator is again well-calibrated for most relevant outcomes: the number of transplantations in primary and repeat kidney transplantation candidates, and the number of transplantations by HLA match quality and immunization status. The simulator does overestimate the number of kidneys transplanted by allocation mechanism in ESP, with on average 23\% too many kidneys placed via non-standard allocation. An apparent consequence of this is that the number of kidneys allocated to candidates aged below 65 is overestimated ($\Plus$28\%), particularly in Belgium, the Netherlands, and Slovenia (see Supplementary Table \ref{tab:esp}) where centers appear to be reluctant to transplant a candidate aged under 65 with an ESP donor. Finally, the simulator is well-calibrated for the number of transplantations by recipient country and match geography, with the only exception Germany where 2\% too many ESP kidneys are transplanted.
\par
Overall, the ETKidney simulator appears to be well-calibrated for most outcomes of ETKAS and ESP allocation. The results of this input-output validation exercise were discussed with medical doctors from Eurotransplant and ETKAC, who deemed differences small enough to make the simulator useful for determining the impact of alternative kidney allocation policies. We illustrate this with case studies in the next section. 
\begin{table}[h]
	\caption{Validation of the number of transplantations between 01-04-2021 and 01-01-2024. For simulations, shown numbers are averages and 95\%-IQRs over 200 simulations. 
		Statistics are displayed in bold if the actual statistic does not fall within the 95\%-IQR.}
	\hspace*{-.03\linewidth}
\centering
\setlength{\tabcolsep}{3pt}
\begin{tabular}{L{3.4cm} R{4cm} L{.9cm} R{4.1cm} L{.9cm}}
\toprule
\multicolumn{1}{c}{ } & \multicolumn{2}{c}{ETKAS} & \multicolumn{2}{c}{ESP} \\
\cmidrule(l{3pt}r{3pt}){2-3} \cmidrule(l{3pt}r{3pt}){4-5}
\multirow[b]{1.5}{*}{category} & \multirow[b]{1.5}{*}{\makecell{simulated results\\(average and 95\%-IQR)}} & \multirow[b]{1.5}{*}{actual} & \multirow[b]{1.5}{*}{\makecell{simulated results\\(average and 95\%-IQR)}} & \multirow[b]{1.5}{*}{actual}\\
& & & & \vspace{-0em} \\

\midrule
\addlinespace[0.3em]
\multicolumn{5}{l}{\textbf{allocation mechanism}}\\
\hspace{.7em}standard & 5044   [4881-5169] & 5000 & \textbf{1117   [-10\%, 1049-1188]} & 1237\\
\hspace{.7em}non-standard & 761    [632-915] & 804 & \textbf{623    [\Plus23\%, 551-690]} & 508\\
\addlinespace[0.3em]
\multicolumn{5}{l}{\textbf{recipient characteristics}}\\
\hspace{.7em}pediatric recipient & 339    [321-358] & 325 & 0 & 0\\
\hspace{.7em}aged 65 or over & 734    [700-769] & 725 & \textbf{1512   [-4\%, 1475-1548]} & 1567\\
\hspace{.7em}aged below 65 & 5071   [5034-5105] & 5079 & \textbf{228    [\Plus28\%, 192-261]} & 178\\
\hspace{.7em}primary transplant & 5038   [5001-5068] & 5039 & 1661   [1644-1677] & 1665\\
\hspace{.7em}repeat transplant & 768    [734-795] & 765 & 80     [67-92] & 80\\
\addlinespace[0.3em]
\multicolumn{5}{l}{\textbf{HLA-ABDR mismatches}}\\
\hspace{.7em}0 ABDR & \textbf{650    [\Plus5\%, 624-675]} & 617 & 3      [1-6] & 4\\
\hspace{.7em}0 or 1 BDR & 1136   [1083-1179] & 1153 & 62     [50-77] & 55\\
\hspace{.7em}1B\Plus1DR or 2B\Plus0DR & 2468   [2406-2534] & 2492 & 316    [285-348] & 336\\
\hspace{.7em}2DR or 3\Plus BDR & 1551   [1490-1609] & 1542 & 1360   [1328-1388] & 1350\\
\addlinespace[0.3em]
\multicolumn{5}{l}{\textbf{sensitization status}}\\
\hspace{.7em}0\% & \textbf{4211   [\Plus2\%, 4173-4255]} & 4142 & 1477   [1455-1502] & 1468\\
\hspace{.7em}$>$0-84.9\% & 1320   [1279-1356] & 1355 & 256    [233-278] & 264\\
\hspace{.7em}85-94.9\% & 173    [155-191] & 186 & 6      [2-10] & 9\\
\hspace{.7em}95\Plus\% & \textbf{101    [-17\%, 87-114]} & 121 & \textbf{2      [-50\%, 1-3]} & 4\\
\addlinespace[0.3em]
\multicolumn{5}{l}{\textbf{recipient country}}\\
\hspace{.7em}Austria & 532    [525-538] & 530 & 106    [87-128] & 118\\
\hspace{.7em}Belgium & 912    [905-918] & 914 & 107    [83-132] & 120\\
\hspace{.7em}Croatia & \textbf{277    [\Plus3\%, 272-283]} & 270 & 38     [26-51] & 26\\
\hspace{.7em}Germany & 2648   [2635-2659] & 2651 & \textbf{1081   [\Plus3\%, 1049-1118]} & 1048\\
\hspace{.7em}Hungary & \textbf{464    [-2\%, 457-470]} & 472 & 19     [11-28] & 24\\
\hspace{.7em}Netherlands & 854    [847-860] & 849 & 380    [349-406] & 398\\
\hspace{.7em}Slovenia & 119    [113-124] & 118 & 9      [4-15] & 11\\
\addlinespace[0.3em]
\multicolumn{5}{l}{\textbf{match geography}}\\
\hspace{.7em}local/regional & \textbf{4087   [\Plus5\%, 3993-4164]} & 3902 & 1374   [1324-1424] & 1408\\
\hspace{.7em}interregional & \textbf{637    [-11\%, 590-680]} & 715 & 174    [138-206] & 178\\
\hspace{.7em}international & \textbf{1081   [-9\%, 1012-1164]} & 1187 & 192    [153-225] & 159\\
\bottomrule
\end{tabular}

	\label{tab:transplant_validation}
\end{table}
\FloatBarrier

\FloatBarrier



\section{Case studies: the impact of modifying ETKAS allocation rules}
\label{sec:case_studies}
Together with medical doctors from Eurotransplant and ETKAC, we selected three topics for case studies in which the ETKidney simulator could help quantify the impact of alternative kidney allocation rules. The selected topics are HLA-BDR matching, introduction of a sliding scale based on the vPRA, and candidate-donor age matching. To limit the effects of transient effects, we extend the simulation period for these case studies from 01-01-2016 to 01-01-2024. We simulate all policy alternatives 20 times, and use traditional hypothesis testing to assess whether alternative policies significantly change the outcomes compared to the current policy. To increase power of these tests, we use common random numbers \cite[~p.588]{lawSimulationModelingAnalysis2015} as a variance reduction technique. Consequently, outcomes can be compared with pairwise t-tests.

\subsection{Case study 1: emphasizing HLA matching on B and DR}
\label{sec:application_dr_matching}
The ETKAS point system has placed equal emphasis on HLA-A, -B and -DR matching since the initiation of ETKAS in 1996, despite common acceptance that HLA-DR mismatches are more deleterious to graft survival than HLA-A and HLA-B mismatches \cite{vereerstraetenExperienceWujciakOpelzAllocation1998,Roberts2004}. Internal analyses on registry data from Eurotransplant suggest that mismatches on the B and DR locus are more strongly associated with graft loss than mismatches on the A locus. This motivated us to simulate policies which emphasize matching on the DR and B loci relative to the A locus. 
\par 
The current ETKAS point system awards 400 points for HLA matching, and penalizes HLA mismatches on the A, B and DR loci with 66.7 points per mismatch. We assess the impact of three alternative HLA matching policies, which all continue to award up to 400 points for candidate-donor HLA match quality but which shift weight from the A locus to the DR and B locus (see Table \ref{tab:alternative_hla_policies}). The first policy is referred to as the $\text{B} + 2\text{DR}$ policy, because it gives no weight to the A locus, maintains the same weight for the B locus (-66.7 points), and doubles the weight on the DR locus (-133.3 points). The second policy, referred to as the $0.5\text{A} + \text{B} + 1.5\text{DR}$ policy, shifts only half of the weight placed on the A locus to the DR locus. The final policy, referred to as $1.5\texttt{B} + 1.5\texttt{DR}$, penalizes mismatches at the B and DR loci both with 100 points per mismatch.
\begin{table}[h!]
	\centering
	\caption{Policy alternatives evaluated for HLA-ABDR matching}
		\begin{tabular}{@{}cccc@{}}
		\toprule
		Policy & $\beta_{\text{MMB}_{HLA_A}}$ &
		$\beta_{\text{MMB}_{HLA_B}}$ & $\beta_{\text{MMS}_{HLA_{DR}}}$ \\
		\midrule
		Current & -66.7     & -66.7     & -66.7\\
		B + 2DR & 0     & -66.7     & -133.3 \\
		$0.5$A + B + $1.5$DR & -33.3 & -66.7     & -100 \\
		$1.5$B + $1.5$DR & 0     & -100     & -100\\
		\bottomrule
	\end{tabular}

	\label{tab:alternative_hla_policies}
\end{table}
\par
Simulation results for these three alternative policies are summarized in Table \ref{tab:results_dr_matching}. When grouping mismatches by a definition of HLA-ABDR match quality used for allocation in the United Kingdom, one can see that the policy alternatives reduces the number of transplantations with 2 DR or 3 or more BDR mismatches by 25 to 39\%, and increase the number of transplantations with 1 B or 1 DR mismatch by 26 to 49\%. Thus, the policies indeed succeed in improving match quality on the B and DR locus.
\par 
However, results also show that there are unintended consequences of this policy: the total number of ABDR mismatches at transplantation increases with all policies, and there are 5 to 12\% fewer transplantations in candidates who are homozygous at the B and/or DR loci, who are already disadvantaged in the current ETKAS system. Results such as those presented in Table \ref{tab:results_dr_matching} can facilitate discussions by ETKAC on whether the improved BDR match quality is worth the increase in total mismatches and reduced access to transplantation for homozygotes. 

\begin{table}[h!]
	\centering
	
	\caption{Simulated change in number of transplantations under alternative HLA-ABDR matching policies. Numbers displayed are the averages difference over 20 simulations. Pairwise t-tests were used to test whether changes in outcomes were statistically significant. mm: mismatches}
	\setlength{\tabcolsep}{5pt}
	
\begin{tabular}{llccc}
\toprule
\multicolumn{2}{c}{ } & \multicolumn{3}{c}{Change in number of transplantations compared to current policy} \\
\cmidrule(l{3pt}r{3pt}){3-5}
  & Current & B + 2DR & 0.5A + B + 1.5DR  & 1.5B + 1.5DR\\
\midrule
\addlinespace[0.3em]
\multicolumn{5}{l}{\textbf{ABDR mismatch count}}\\
\hspace{.5em}0 & 1925 & -9 & -1 & -6\\
\hspace{.5em}1 & 952 & -181*** & -55*** & -99***\\
\hspace{.5em}2 & 3855 & -634*** & -252*** & -509***\\
\hspace{.5em}3 & 6037 & -358*** & -76*** & -224***\\
\hspace{.5em}4 & 3235 & \Plus 837*** & \Plus 282*** & \Plus 642***\\
\hspace{.5em}5 & 730 & \Plus 340*** & \Plus 104*** & \Plus 191***\\
\hspace{.5em}6 & 78 & \Plus 2 & 0 & \Plus 7*\\
\addlinespace[0.3em]
\multicolumn{5}{l}{\textbf{ABDR match quality}}\\
\hspace{.5em}000 & 1925 & -9 & -1 & -6\\
\hspace{.5em}*00, *10, *01 & 3264 & \Plus 1364*** & \Plus 850*** & \Plus 1585***\\
\hspace{.5em}*20, *11 & 7234 & \Plus 209*** & \Plus 240*** & \Plus 131***\\
\hspace{.5em}**2, *21 & 4388 & -1567*** & -1087*** & -1708***\\
\addlinespace[0.3em]
\multicolumn{5}{l}{\textbf{candidate homozygosity on B and DR}}\\
\hspace{.5em}B and DR & 473 & -39*** & -17*** & -30***\\
\hspace{.5em}DR & 1662 & -207*** & -104*** & -94***\\
\hspace{.5em}B & 1052 & \Plus 31*** & \Plus 14** & -19***\\
\hspace{.5em}none & 13625 & \Plus 212*** & \Plus 110*** & \Plus 144***\\
\bottomrule
\end{tabular}

	\label{tab:results_dr_matching}
	\\[.1cm]
	\footnotesize{$\qquad\qquad^{*}p<0.05; \ ^{**}p<0.01; \  ^{***}p<0.001$\hfill}
\end{table}
\par

\FloatBarrier

\subsection{Case study 2: a sliding scale for the vPRA}
\label{sec:application_vpra}
Candidates with unacceptable antigens indirectly receive priority in ETKAS through mismatch probability points (MMPPs), see Section \ref{subsection:etkas}. Despite this form of compensation, studies on Eurotransplant registry data have shown that immunized candidates face longer waiting times than non-immunized candidates in ETKAS \cite{ziemannUnacceptableHumanLeucocyte2017, zecherImpactSensitizationWaiting2022a, deferranteImmunizedPatientsFace2023}. In this case study, we assess whether this disparity can be alleviated by awarding points directly for the vPRA.
\par 
For this, we modify the ETKAS point system in two ways. The first modification is that we directly award points for the vPRA using a \textit{sliding scale}. Such a sliding scale for the vPRA has been a part of kidney allocation in the United States since 2014 \cite{stewartSmoothingItOut2012}. This sliding scale is parameterized by a weight, $\beta_{\texttt{vPRA}}$, which determines the maximum number of points awarded for the vPRA, and a base $b$, which controls the steepness of the sliding scale. 
The second modification is that we no longer directly award points for the vPRA via the mismatch probability. Instead, we replace mismatch probability points by \textit{\enquote{HLA mismatch probability points}} (HMPPs) (Section \ref{subsection:mmp}). These HMPPs are calculated as $\texttt{HMPP} = \Big[1-f_{\leq1mm}\Big]^{1000}$, where $f_{\leq1mm}$ is the 1-ABDR HLA mismatch frequency (see Section \ref{subsection:mmp}). For calculation of HMPPs, we do not use the candidate blood group, which is motivated by the fact that ETKAS allocation has been blood group identical since 2010.
\par
\begin{figure}[h]
	\centering
	\includegraphics[width=0.8\linewidth]{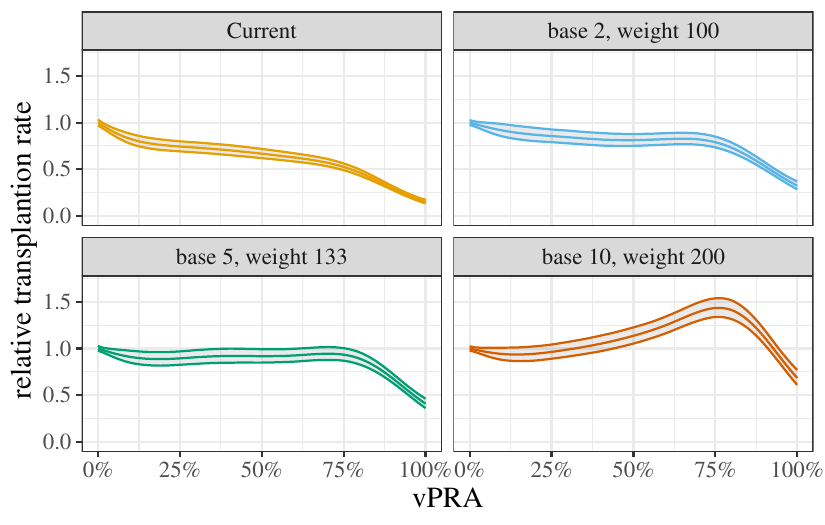}
	\caption{Relations between the relative transplantation rate and vPRA in ETKAS, estimated on ETKidney simulator outcomes. These relations were estimated with a Cox proportional hazards model, adjusting for the vPRA using spline terms. }
	\label{fig:vpraeffect}
\end{figure}

In discussing this case study, representatives of ETKAC reached consensus that the aim of a sliding scale should be that a candidate's chance to be transplanted through ETKAS should not decline up to a vPRA of 85\%. Above this vPRA, candidates could have access to the AM program, or should consider removing unacceptable antigens in case they do not meet AM criteria. We used the ETKidney simulator to simulate ETKAS for different combinations of weights $\beta_{\texttt{vPRA}}$ and bases $b$, and quantify the association between vPRA and the relative transplantation rate using Cox proportional hazards model on simulated outcomes. For this, the same model specification is used as in a previous analysis of the association between vPRA and the transplantation rate \cite{deferranteImmunizedPatientsFace2023}. In Figure \ref{fig:vpraeffect} the estimated relation between vPRA and the transplantation rate is shown for several sliding scales. From this figure, the most acceptable option would be a sliding scale with a base of 5 and weight of 133, with which the relative transplantation rate no longer decays until a vPRA of 85\%.

\subsection{Case study 3: candidate-donor age matching}
\label{sec:application_age_matching}
Consensus in the transplantation literature is that kidneys procured from young donors should preferentially be transplanted in young candidates \cite{waiserAgeMatchingRenal2000, pippiasYoungDeceasedDonor2020, vanittersumIncreasedRiskGraft2017, coemansCompetingRisksModel2024, keithEffectDonorRecipient2004b}. While allocation systems in France and the United Kingdom prioritize such candidate-donor age matching \cite{audryNewFrenchKidney2022a, watsonOverviewEvolutionUK2020}, the \WOP\ does not award points based on candidate or donor age (apart from pediatric points). In this case study, we (i) quantify the associations of candidate and donor age with graft and patient survival retrospectively on Eurotransplant registry data using cause-specific hazard models, and (ii) simulate and evaluate two age matching policies for ETKAS.
\par
\begin{figure}[h]
	\centering
	\includegraphics[width=\linewidth]{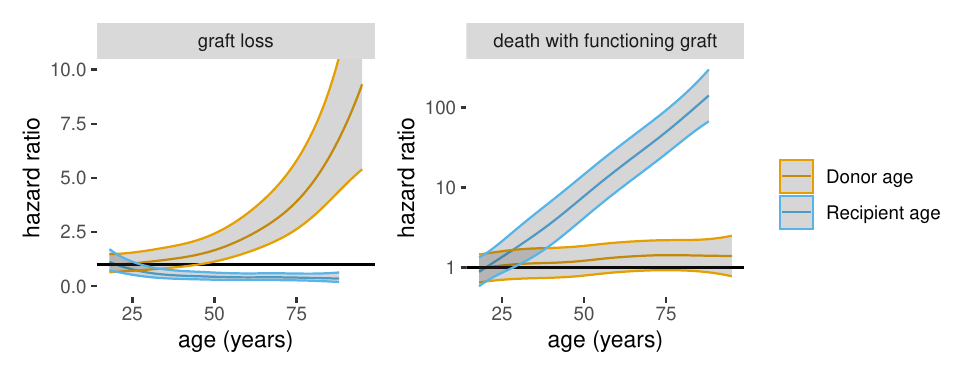}
	\caption{Estimated relations between graft loss and death with a functioning graft. Note that the hazard ratio for death with functioning graft is shown on the logarithmic scale.}
	\label{fig:competingriskmodel}
\end{figure}

To quantify the relation between donor or candidate age and post-transplant survival, we consider all patients transplanted with a kidney through ETKAS or ESP between 2004 and 2019. We exclude candidates with the HU status and candidates without any follow-up information ($n = 7,458$), leaving $n = 36,576$ transplantations. We fit cause-specific Cox proportional hazards models on these transplantations for (i) graft loss and (ii) death with a functioning graft. We censored both time-to-event variables ten years after transplantation, because completeness of follow-up data for more than 10 years is poor in the Eurotransplant registry. Besides donor and candidate age, we adjust for donor characteristics (heart or non-heartbeating donation, hypertension, last creatinine, death cause, diabetes, malignancy), candidate characteristics (dialysis time), and match characteristics (zero mismatch, number of mismatches per locus for the A, B, and DR loci, match geography). To allow for non-linear relations between continuous variables and the hazard rate, we adjust for spline transformations of the continuous variables. The estimated relations between donor / candidate age and patient / graft survival are shown in Figure \ref{fig:competingriskmodel}. These results are qualitatively similar to results by Coemans et al. \cite{coemansCompetingRisksModel2024}, who report that the hazard rate of graft loss decays linearly with recipient age while it increases quadratically with donor age, and that the mortality hazard rate increases quadratically with candidate age. 
\par
\par
We simulate two candidate-donor age matching policies, which were both inspired by the 2015 French kidney allocation policy \cite{audryNewFrenchKidney2022a}. Like ETKAS, the French policy awards points for candidate waiting time, HLA match, the candidate's likelihood to be favorably matched with a kidney, and the geographic distance between the donor and candidate. However, in France an \enquote{age filter} is applied to the total number of points, with candidates ranked based on the filtered number of points. For instance, the French age filter is 0\% for a candidate who is 20 or more years older than the donor, which means that such candidates receive 0\% of the total points for ranking. This French age filter is asymmetrical, with allocation of kidneys from a young donor in an older patient discouraged more strongly than allocation of a kidney from an older donor to a young candidate.
\par
\begin{figure}[h]
	\centering
	\includegraphics[width=0.85\linewidth]{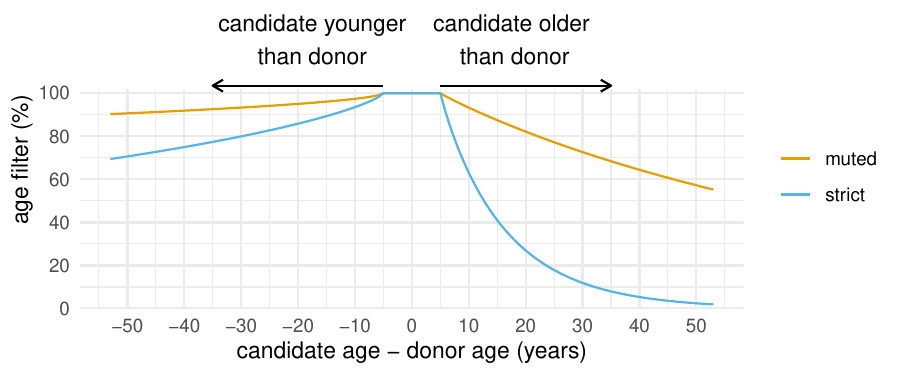}
	\caption{Evaluated age filters for ETKAS.}
	\label{fig:baseagefiltermuted}
\end{figure}

We evaluate two scenarios for such an asymmetrical age filter for ETKAS (see Figure \ref{fig:baseagefiltermuted}). Both filters give a candidate 100\% of their \WO\ points in case the difference in candidate age and donor age is 5 years or less. The \enquote{strict} filter (blue) is similar to the French age filter in that it gives almost no points in case the candidate is much older than the donor. The \enquote{muted} filter (orange)  gives a larger fraction of the total number of \WO\ points, which we anticipated to be more acceptable for Eurotransplant because it maintains a better balance in the international exchange of kidneys.
\FloatBarrier
Simulated outcomes for the muted and strict age matching policies are compared to the current policy in Table \ref{tab:results_age_matching}. The Table shows that the muted and strict policies increase age-matched transplantations (maximum 5 year difference in donor and candidate age) by 60\% and 138\%, respectively. An unintended consequence is that such age matching leads to reduced HLA match quality, with a 12 and 33\% increase in level 4 mismatched kidney transplantations (2 DR, or 3 or 4 B+DR mismatches). In terms of match geography (third panel), Table \ref{tab:results_age_matching} shows that the muted policy modestly increases international sharing (+5\%) and inter-regional sharing (+7\%), while the strict policy leads to a 37\% increase in international transplantations. Simulation results also show that the strict age filter also increases the number of extended or rescue transplantations.
\par
\begin{figure}[h]
	\centering
	\includegraphics[width=0.6\linewidth]{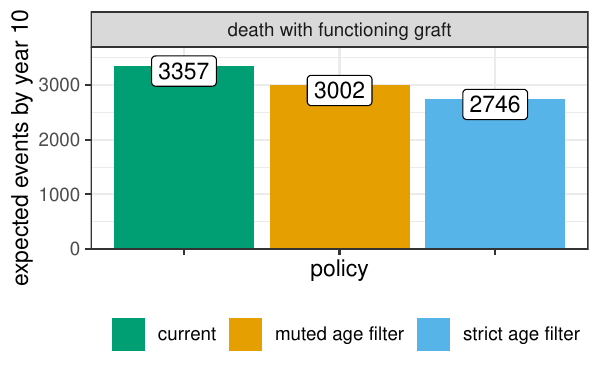}
	\caption{Expected number of post-transplant events 10 years after transplantation, predicted based on candidate, donor, and transplantation characteristics with competing risk models.}
	\label{fig:posttxpeventsagematching}
\end{figure}

 
To assess whether improvements in candidate-donor age matching outweigh the unintended consequences, we use the earlier mentioned cause-specific hazard models to predict the probability of death with a functioning graft ten years after transplantation for all simulated transplantations. For this, we use a competing risk approach \cite{wreedeMstatePackageAnalysis2011}. The expected number of events ten years after transplantation is visualized in Figure \ref{fig:posttxpeventsagematching} for the current policy (green), the muted age filter (orange), and the strict age filter (blue). These results suggest that the muted and strict age filter could reduce the numbers of deaths with a functioning graft 10 years after transplantation by 11\% and 18\%, respectively.

\begin{table}[h!]
	\centering
	\caption{Simulated change in the number of transplantations in ETKAS with the age matching policies. Numbers displayed are the averages difference over 20 simulations. Pairwise t-tests were used to test whether changes in outcomes were statistically significant. mm: mismatches.}
	
\begin{tabular}{llcc}
\toprule
\multicolumn{2}{c}{ } & \multicolumn{2}{c}{\multirow[b]{1.5}{*}{\makecell{Change in number of transplantations\\compared to current policy}}} \\
& & & \vspace{-0em} \\
\cmidrule(l{3pt}r{3pt}){3-4}
  & current & muted age filter & strict age filter\\
\midrule
\addlinespace[0.3em]
\multicolumn{4}{l}{\textbf{age difference}}\\
\hspace{1em}candidate 35+ years older & 708 & -497$^{***}$ & -557$^{***}$\\
\hspace{1em}candidate 15-34 years older & 3166 & -1742$^{***}$ & -2549$^{***}$\\
\hspace{1em}candidate 6-14 years older & 3256 & +434$^{***}$ & -1756$^{***}$\\
\hspace{1em}max 5 year difference & 4810 & +2919$^{***}$ & +6640$^{***}$\\
\hspace{1em}candidate 6-14 years younger & 2624 & -151$^{***}$ & -230$^{***}$\\
\hspace{1em}candidate 15-34 years younger & 2064 & -873$^{***}$ & -1386$^{***}$\\
\hspace{1em}candidate 35+ years younger & 184 & -86$^{***}$ & -155$^{***}$\\
\addlinespace[0.3em]
\multicolumn{4}{l}{\textbf{HLA match quality}}\\
\hspace{1em}level 1 (0 ABDR mm) & 1922 & -8 & -40$^{***}$\\
\hspace{1em}level 2 (at most 1 BDR mm) & 3250 & -380$^{***}$ & -902$^{***}$\\
\hspace{1em}level 3 (2B or 1DR+1B mm) & 7224 & -156$^{***}$ & -496$^{***}$\\
\hspace{1em}level 4 (2DR or 3\Plus BDR mm) & 4417 & +547$^{***}$ & +1446$^{***}$\\
\addlinespace[0.3em]
\multicolumn{4}{l}{\textbf{match geography}}\\
\hspace{1em}local/regional & 11744 & -297$^{***}$ & -1514$^{***}$\\
\hspace{1em}national & 1796 & +121$^{***}$ & +306$^{***}$\\
\hspace{1em}international & 3272 & +179$^{***}$ & +1215$^{***}$\\
\addlinespace[0.3em]
\multicolumn{4}{l}{\textbf{type of allocation}}\\
\hspace{1em}standard allocation & 14487 & +50 & -160*\\
\hspace{1em}non-standard & 2325 & -47 & +167$^*$\\
\bottomrule
\end{tabular}
\\[.1cm]
\footnotesize{$\qquad\qquad^{*}p<0.05; \ ^{**}p<0.01; \  ^{***}p<0.001$\hfill}
	\label{tab:results_age_matching}
\end{table}

\FloatBarrier

\section{Discussion and conclusion}
\label{sec:discussion}
Eurotransplant recognized the importance of computer simulations for allocation development already in the 1998 one-year evaluation of ETKAS, in which it was stressed that  \textit{\enquote{introduction of a change must be preceded by a computer simulation study}} \cite{demeesterNewEurotransplantKidney1998}. However, Eurotransplant has only recently started the development of tools required for such simulation studies; in 2022, a liver simulator was developed which informed updating of Eurotransplant's liver allocation policy \cite{deferranteELASSimulator2023}, and in 2024, the thoracic committee recommended development of a simulation tool for heart and lung allocation. Within this line of research, we present the ETKidney simulator.
\par
Such simulators are used routinely to update allocation rules in other geographic regions \cite{Pritsker1995,Mumford2018,jacquelinetChangingKidneyAllocation2006a}. The most prominent simulator is the Kidney-Pancreas Simulated Allocation Model (KPSAM), which is developed and maintained by the SRTR U.S. kidney allocation, and publicly available for research. KPSAM differs from the ETKidney simulator in several aspects. Firstly, KPSAM users have to manually specify in simulation inputs when a candidate would list for a repeat transplantation, as well as how their status updates would evolve. In the ETKidney simulator, simulation of re-listings is instead based on historical data. Secondly, graft offer acceptance behavior is simulated according to a single, patient-level logistic regression in KPSAM, whereas the ETKidney simulator additionally includes logistic regressions at the center level to capture that centers regularly decline kidneys for all their candidates. Thirdly, available kidneys are discarded in KPSAM after a fixed number of offers, whereas the ETKidney simulator has functionality to model the maximum number of offers made in standard allocation based on donor characteristics. Once this number is reached, the ETKidney simulator can also switch to non-standard allocation, whereas such out-of-sequence offering is not simulated in KPSAM.
\par 
More important than these technical differences is that the ETKidney simulator is a bespoke model for Eurotransplant, which implements Eurotransplant-specific allocation mechanisms such as mismatch probability points and the balance system. Eurotransplant needs such a bespoke model in communication with national competent authorities, who are interested not only in the overall effects of policies, but also in the specific effects that a policy change has on their national waiting list population. 
\par 
To build trust in the simulator, we have used input-output validation to show that the simulator can closely approximate contemporary transplantation patterns of ETKAS and ESP. Results of this validation exercise were discussed with medical doctors from Eurotransplant, and presented at the Eurotransplant Annual Meeting to other major stakeholders, such as nephrologists, immunologists, and transplantation coordinators from the Eurotransplant's kidney transplantation centers, as well as representatives from the national competent authorities. In the views many of these stakeholders, the simulator has become a useful tool for kidney allocation policy development.
\par
We acknowledge that the ETKidney simulator also has limitations. Firstly, the graft offer acceptance and post-transplant survival models are calibrated to historical data. These models may lack external validity for future post-transplant survival and future offer acceptance behavior. An example of this, encountered during input-output validation, is that the simulator appears to underestimate the number of transplantations in immunized candidates after introduction of the virtual crossmatch in January 2023. This limitation could be addressed by refitting the organ acceptance models in the future on more contemporary data. A second limitation validation was only based on historic input-output validation. Ideally, we would have been able to also observe ETKAS and ESP outcomes under alternative sets of allocation rules, and study whether the simulator would be able to capture simulated outcomes under these alternative rules \cite{sargent2020}. However, this was not feasible because ETKAS and ESP allocation rules have changed little since their introductions in 1996 and 1999, respectively. A final limitation is that Eurotransplant is not allowed to publicly release information which could potentially identify its donors and candidates, which precludes exact reproduction of our simulations by external parties. We have tried to address this limitation by making synthetic data available, on which kidney allocation could be simulated.
\par
Overall, we are confident that the ETKidney simulator is a valuable tool for quantifying the impact of kidney allocation policy changes in Eurotransplant, as we demonstrated with three clinical case studies. We anticipate that the simulator can play a pivotal role in modernizing ETKAS and ESP allocation rules, in collaboration with subject-matter experts from ETKAC, ETRL, and national competent authorities.

\newpage

\newpage
\bibliography{refs}
\newpage

\appendix
\renewcommand\thefigure{\thesection\arabic{figure}}
\renewcommand{\thetable}{\Alph{section}\arabic{table}}
\setcounter{figure}{0}
\setcounter{table}{0}

\section{ETKAS and ESP Allocation Rules} 
\label{app:allocation rules}
This appendix describes the eligibility, filtering and ranking criteria of ETKAS and ESP rules introduced in March 2021, and includes examples of ETKAS and ESP match lists.

\FloatBarrier
\subsection{Eligibility, filtering and ranking criteria in ETKAS}
\label{subsection:criteria_etkas}
Eligibility criteria determine whether a candidate can appear on the unfiltered match list. The eligibility criteria used in ETKAS are:
\begin{enumerate}[noitemsep]
	\item the candidate must have the same blood group as the donor,\footnote{Since 2010. Before 2010, transplantations across blood groups were allowed in case the donor and candidate had 0 HLA-ABDR mismatches. This was abandoned because it increased waiting times for blood group O candidates, who can only accept kidneys from blood group O donors.},
	\item the candidate should not have the non-transplantable (NT) status, which is a status used by centers to indicate that their candidate is (temporarily) unavailable for transplantation,
	\item the candidate's HLA typing at the A, B, and DR loci must be known, and the candidate is not allowed to have reported unacceptable antigens against the HLAs present in the donor's HLA typing,
	\item an antibody screening for the candidate must have been reported to Eurotransplant within the last 180 days,
	\item in Germany, candidates above age 65 must have chosen to not be included in ESP\footnote{ETKAS and ESP have been mutually exclusive in Germany since 2010},
	\item the candidate is also not allowed to have an active status in the AM program.
\end{enumerate}
Filtering criteria determine whether Eurotransplant will contact the transplantation center in standard allocation to offer the kidney for transplantation of a named candidate. For ETKAS, used filtering criteria are:
\begin{enumerate}[noitemsep]
	\item the allocation profile, which is used by centers to specify that their candidates does not want to accept certain donor characteristics. Selectable donor characteristics are (i) a minimum and/or maximum donor age, (ii) virology of the donor, such as whether the donor is hepatitis C positive, (iii) whether the donor is a non-heartbeating donor, (iv) extended donor criteria, which include whether the donor has had a malignancy, sepsis, meningitis, euthanasia, or history of drug abuse, and whether the kidney donor is procured from a domino transplantation.
	\item HLA mismatch criteria. Centers can use these to specify that they do not want to consider offers of a certain candidate-donor HLA match qualities. For instance, most centers exclude offers with 6 mismatches in total, and many centers continue to exclude offers with 2B or 2DR mismatches which were used as minimal match criteria in Eurotransplant from 1988 to 1996 \cite{demeesterNewEurotransplantKidney1998}.
\end{enumerate}
\par
An example of a \textit{filtered} ETKAS match list for a Belgian donor is shown in Table \ref{tab:example_etkas_list}. The ranking of candidates on this list is determined by three tiers and the Wuijciak-Opelz point system. The three ETKAS tiers are the zero mismatch tier, the pediatric tier, and the non-zero mismatch-tier \cite{manualKidney}. Candidates in higher tiers have absolute priority over candidates in lower tiers, regardless of their point score. The zero-mismatch tier (0MM-tier) consists of candidates with 0 HLA-ABDR mismatches with the donor. Within the 0MM-tier, subtiers exist to prioritize candidates who are homozygous at the HLA-ABDR loci in case the donor is fully homozygous.\footnote{These subtiers are based on the \textit{homozygosity level} of the candidate. For instance, a candidate with homozygosity on one of the loci would have priority over candidates heterozygous for all loci, regardless of their point score.} The pediatric tier is used to prioritize the allocation of kidneys procured from pediatric donors to pediatric candidates. All other candidates are included in the non-zero mismatch tier ($>$0MM-tier). The Wujciak-Opelz point system is explained in main text.
\par
We point out that many candidates who met the eligibility criteria for this kidney did not appear on the match list because of filtering criteria. For instance, Table \ref{tab:example_etkas_list} shows that the right donor kidney was accepted by the candidate ranked 14th on the filtered match list. In case filtering criteria were not applied, this candidate would have appeared on the 67th position.
\par
\begin{table}[h]
	\hspace*{-.1\linewidth}
	\resizebox{1.2\linewidth}{!}{
		\centering
		
\begin{tabular}{ccccccccccccc}
	\toprule
	\multicolumn{6}{c}{ } & \multicolumn{6}{c}{match point components} & \multicolumn{1}{c}{ } \\
	\cmidrule(l{3pt}r{3pt}){7-12}
	\makecell{tier} & \makecell{listing\\country} & \makecell{ABDR\\match\\quality} & \makecell{time on\\dialysis\\(years)} & \makecell{rank} & \makecell{total\\points} & dialysis & \makecell{HLA\\match} & \makecell{pediatric\\bonus} & \makecell{balance} & \makecell{distance} & \makecell{MMP} & \makecell{accepted}\\
	\midrule
	0MM & Germany & 000 & 9.0 & 1 & 722 & 298 & 400 & 0 & 0 & 0 & 24 & LKi\\
	\cmidrule{1-13}
	& Croatia & 111 & 7.5 & 2 & 1343 & 249 & 400 & 100 & 550 & 0 & 44 & -\\
	\cmidrule{2-13}
	&  & 111 & 4.7 & 3 & 1300 & 155 & 200 & 0 & 550 & 300 & 95 & -\\
	\cmidrule{3-13}
	&  & 111 & 3.0 & 4 & 1219 & 100 & 200 & 0 & 550 & 300 & 69 & -\\
	\cmidrule{3-13}
	& \multirow{-3}{*}{\centering\arraybackslash Belgium} & 202 & 2.7 & 5 & 1156 & 90 & 133 & 0 & 550 & 300 & 83 & -\\
	\cmidrule{2-13}
	& Hungary & 001 & 0.0 & 6 & 1147 & 0 & 667 & 100 & 370 & 0 & 10 & -\\
	\cmidrule{2-13}
	&  & 101 & 0.2 & 7 & 1143 & 7 & 267 & 0 & 550 & 300 & 19 & -\\
	\cmidrule{3-13}
	&  & 102 & 0.0 & 8 & 1130 & 0 & 200 & 0 & 550 & 300 & 80 & -\\
	\cmidrule{3-13}
	&  & 101 & 5.2 & 9 & 1113 & 172 & 267 & 0 & 550 & 100 & 24 & -\\
	\cmidrule{3-13}
	&  & 111 & 0.1 & 10 & 1097 & 5 & 200 & 0 & 550 & 300 & 42 & -\\
	\cmidrule{3-13}
	&  & 111 & 0.0 & 11 & 1066 & 0 & 200 & 0 & 550 & 300 & 16 & -\\
	\cmidrule{3-13}
	&  & 110 & 1.8 & 12 & 1059 & 60 & 267 & 0 & 550 & 100 & 82 & -\\
	\cmidrule{3-13}
	&  & 202 & 0.7 & 13 & 1053 & 24 & 133 & 0 & 550 & 300 & 46 & -\\
	\cmidrule{3-13}
	\multirow{-13}{*}[1.5\dimexpr\aboverulesep+\belowrulesep+\cmidrulewidth]{\centering\arraybackslash $>$0MM} & \multirow{-8}{*}{\centering\arraybackslash Belgium} & 110 & 1.4 & 14 & 1049 & 47 & 267 & 0 & 550 & 100 & 85 & RKi\\
	\bottomrule
\end{tabular}

	}
	
	\caption{Example of a filtered ETKAS match list for a blood group A donor reported in Belgium. The donor's left kidney was accepted on rank 1, the right kidney on rank 14. One candidate has 0 HLA-ABDR mismatches with the donor, and is ranked on in the 0MM-tier. Within tiers, priority is determined by the Wujciak-Opelz point system. Only candidates registered in the same region as the donor receive 300 distance points in Belgium.
	}
	\label{tab:example_etkas_list}
\end{table}

\FloatBarrier
\newpage
\subsubsection{Eligibility, filtering and prioritization in the ESP program}
\label{subsection:esp_program}
Eligibility criteria used in ESP are:
\begin{enumerate}[noitemsep]
	\item candidates must have an active waiting list status,
	\item the candidate must be aged over 65, or have specified that they want to receive offers through extended ESP allocation in case they are aged under 64,
	\item the candidate has to be blood group identical with the donor,
	\item the candidate must have reported a valid HLA typing, and a valid antibody screening within the last 180 days,
	\item candidates who report unacceptable antigens are only eligible for ESP if the donor HLA is known, and no unacceptable antigens are present in the donor HLA,
	\item candidates cannot have an active AM status.
\end{enumerate}
International candidates are eligible for offers via ESP since March 2021. 
\par
Filtering criteria used for ESP are:
\begin{enumerate}[noitemsep]
	\item the candidate is aged below 65,
	\item the candidate is a German candidate who has chosen for the ETKAS program,
	\item the donor has to be compatible with the candidate's allocation profile.
\end{enumerate}
We point out that HLA mismatch criteria are not used as a filtering criterion in ESP.
\par
Tiers are used to rank candidates in ESP. These tiers are based on the candidate's geographical proximity to the donor, and they differ per ET member country. For instance, in Germany ESP offers are first made to candidates who are located in the same ESP subregion as the donor, and then to candidates located in the same region, while in the Netherlands all candidates appear in a single tier (see the ET manual \cite{manualKidney}). In March 2021, subtiers were introduced in ESP for candidates with a High Urgency (HU) status and the Kidney After Other Organ (KAOO) status. Within tiers, candidates are ranked by the number of days they have waited on dialysis.
\par 
Table \ref{tab:example_esp_match_list} shows an example of a filtered ESP match list for a donor reported from Germany. The right and left kidney of this donor were accepted at rank 2 and 11, respectively. Both offers were accepted by candidates located in the ESP subregion consisting of Stuttgart, Tübingen, Mannheim, and Heidelberg (the highest tier for this donor). The order of candidates within a tier is based on the time spent on dialysis.

\begin{table}[h]
	\resizebox{\linewidth}{!}{
		\centering
		
\begin{tabular}{ccccccc}
	\toprule
	listing country & donor region & listing center & time on dialysis (days) & rank & total points & accepted\\
	\midrule
	&  & Stuttgart & 1143 & 1 & 1143 & -\\
	\cmidrule{3-7}
	&  & Tübingen & 964 & 2 & 964 & RKi\\
	\cmidrule{3-7}
	&  & Heidelberg & 890 & 3 & 890 & -\\
	\cmidrule{3-7}
	&  & Tübingen & 871 & 4 & 871 & -\\
	\cmidrule{3-7}
	&  & Stuttgart & 867 & 5 & 867 & -\\
	\cmidrule{3-7}
	&  & Tübingen & 855 & 6 & 855 & -\\
	\cmidrule{3-7}
	&  &  & 715 & 7 & 715 & -\\
	\cmidrule{4-7}
	&  & \multirow{-2}{*}{\centering\arraybackslash Heidelberg} & 714 & 8 & 714 & -\\
	\cmidrule{3-7}
	&  & Tübingen & 596 & 9 & 596 & -\\
	\cmidrule{3-7}
	&  &  & 423 & 10 & 423 & -\\
	\cmidrule{4-7}
	\multirow{-11}{*}[4\dimexpr\aboverulesep+\belowrulesep+\cmidrulewidth]{\centering\arraybackslash Germany} & \multirow{-11}{*}[4\dimexpr\aboverulesep+\belowrulesep+\cmidrulewidth]{\centering\arraybackslash \makecell{Baden-\\Württemberg}} & \multirow{-2}{*}{\centering\arraybackslash Mannheim} & 419 & 11 & 419 & LKi\\
	\bottomrule
\end{tabular}

	}
	\caption{Example of an ESP match list for a blood type O donor reported in Baden-Württemberg. ESP donors are only offered to candidates located in vicinity of the donor, in this case Stuttgart, Tübingen, Mannheim, and Heidelberg. Candidates are currently solely ranked by their accrued dialysis time.
	}
	\label{tab:example_esp_match_list}
\end{table}

\FloatBarrier
\subsubsection{Deviation from the standard allocation procedure}
\label{subsection:rescue}
To avoid kidney nonuse, Eurotransplant deviates under specific conditions from standard allocation with extended or rescue allocation. These conditions include:
\begin{itemize}[noitemsep]
	\itemsep0em
	\item when a kidney has been turned down for donor or quality reasons by five different centers,
	\item when an ESP donor has not been allocated five hours after it was procured,
	\item when all candidates on the filtered ESP match list have received an offer (since March 2021),
	\item when loss of a transplantable graft is anticipated because of the cancellation of a planned transplantation procedure.
\end{itemize}
In general, Eurotransplant first tries to place the kidney via extended allocation, and resorts to rescue allocation only if loss of a transplantable kidney is anticipated.
\par 
Extended allocation was implemented in December 2013, and in this procedure centers located in proximity of the kidney are contacted in parallel by phone for the kidney offer. Via an online application, these centers can see which candidates meet the eligibility criteria for the kidney. Centers can then select two candidates for transplantation from this list, and the candidate who would have achieved the highest rank on the original match list receives the kidney offer. In general, Eurotransplant first tries to place the kidney via extended allocation, and resorts to rescue allocation only if loss of a transplantable kidney is anticipated.
\par
In case extended allocation is unsuccessful, or in case of an imminent loss of a transplantable kidney, Eurotransplant starts the rescue allocation procedure. In this procedure, at least three centers located in proximity of the donor are contacted by phone. The first center to propose a candidate for transplantation to Eurotransplant receives the offer. This candidate does not have to meet ETKAS eligibility criteria.

\section{Details to the future status update imputation procedure}
\label{app:imp_proc}
\setcounter{table}{0}
The simulations in this paper use historic candidate information from the Eurotransplant registry for the simulation of ETKAS and ESP allocation. A challenge in using historic candidate information is that transplantation of a waiting list candidate prevents Eurotransplant from observing what would have happened to this waiting list candidate in case they had remained on the waiting list. To realistically complete the waiting list spells for candidates whom were transplanted with status updates until they would have experienced a waiting list death or waiting list removal, we implemented a procedure to impute future statuses for transplant recipients. This appendix details this procedure.

\bibliographystyleAC{myama}

\paragraph*{Overview of the \textit{counterfactual} future status imputation procedure}
The ETKidney simulator requires a complete stream of status updates for all patient registrations, i.e. all registrations must end with a removal (status code: R) status or a waiting list death (status code: D) status. In registry data from Eurotransplant, most waiting list spells end in a transplantation (status code: FU). For this, transplant recipients are matched to candidates who remain on the waiting list based on their expected  \textit{counterfactual} mortality risk as well as on other characteristics. In predicting this counterfactual mortality, we correct for dependent censoring by transplantation with inverse probability censoring weighting by modifying a procedure proposed by Tayob and Murray \citeAC{tayobstatistical2017} to impute future survival times from pre-determined landmark times for lung waiting list transplantation candidates.
\par
Specifically, Tayob and Murray are interested in modelling the 12-month restricted survival time $T^* = \min(T, \tau)$ of transplant candidates, where $T$ is a random variable for the remaining survival time measured from the landmark times, and $\tau$ is the time horizon of interest (12 months in their case). Tayob and Murray work with regularly spaced landmark times $j$, and define $T^*_{ij}$ as the $\tau$-restricted remaining survival time of subject $i$ from landmark time $j$. With these, Tayob and Murray aim to model the expected mean survival time directly, i.e. model
$$\mathbb{E}[\log(T^*) | Z] = \beta^\intercal Z.$$ There are three challenges in estmiating $\beta$: (i) $T^*_{ij}$ is unobserved for most candidates due to censoring, (ii) transplantation represents an informative censoring mechanism, and (iii) that the $T^*_{ij}$ exhibit correlations over the landmark times $j$.
\par
The paper by Tayob and Murray proposes a procedure to handle these three issues. Specifically, Tayob and Murray's imputation procedure consists of:
\begin{enumerate}
	\item Estimating the counterfactual waiting list survival function $S^{\text{IPCW}}_T(t)$, with correction for informative censoring by transplantation with inverse probability censoring weighting (IPCW). With this curve, they can construct $j$-specific pseudo-observations $PO_{ij}$ for $\log(T^*_{ij})$. 
	\item Estimating $\hat{\beta}^{\text{PO}}$ on pairs $(PO_{ij}, Z_{ij})$ with Generalized  Estimating Equations (GEE) with an unstructured working correlation matrix. This correlation matrix permits arbitrary correlation between the $T_{ij}$s over $j$.
	\item For each patient $i$ with unobserved $T^*_{ij}$ constructing a risk set $R_i$
	\begin{itemize}
		\item of patients who remain at risk after patient $i$ is censored, i.e. with $T_{kj} > C_{ij}$, $k\neq i$, where $C_{ij}$ is patient $i$'s censoring time measured from landmark time $j$,
		\item who are similar in terms of predicted expected log survival, i.e. require $$|\hat{\beta}^{\text{PO}}\ ^\intercal Z_k(C_{i}) - \hat{\beta}^{\text{PO}}\ ^\intercal Z_i(C_{i})| < \epsilon$$ for some $\epsilon$, where $Z_k(C_{i})$ are covariates of patient $k$ at $i$'s censoring time $C_i$,
		\item Optionally, further restrict matches based on $Z_i(C_{i})$, e.g. require matching disease groups.
	\end{itemize}
	\item Within risk set $R_i$, re-estimate the counterfactual survival function $S_T^{\text{IPCW}}(t|R_i)$. Use inverse transform sampling from this risk-set specific survival curve to sample a valid $T^*_{ij}.$
\end{enumerate}
\par 
Our aim is to use Tayob and Murray's procedure to match transplant recipients to comparable, not-yet-transplanted waiting list candidates. Because priority is based on dialysis time in Eurotransplant, we want to match candidates based on accrued dialysis time. This requires three main modifications to the imputation procedure:
\begin{itemize}
	\item We define the remaining time-to-event variables based on accrued dialysis time instead of waiting time,
	\item We have to match transplanted patients at any continuous time $t$ to comparable, at-risk waiting list candidates (and not only at pre-determined landmark times $j$). For this, we define $T^*_{it}$ to be the $\tau$-restricted remaining survival time from time $t$ forward, with $t$ the number of days elapsed since listing. For each observed status update we construct a $t$-specific pseudo-observation $PO_{it}$, with $t$ the time-after-listing at which the status was reported.
	\item We cannot estimate ${\beta}^{\text{PO}}$ with GEE with an unstructured correlation matrix, as the $PO_{it}$ is indexed by continuous time $t$. Instead, we propose to estimate $\beta^{PO}$ with Quasi-Least Squares (QLS) with a Markov correlation structure \citeAC{xieqls2010}. This structure assumes that the correlation between $PO_{it}$ decays with spacing in $t$. 
\end{itemize} 
Construction of the risk set $R_i$ for patients with unknown $T^*_{it}$ is then similar; we consider patients $k$ with $T^*_{kt} > C^*_{it}$, similar $\hat{\beta}^{\text{PO}}\ ^\intercal Z_\cdot(C_{ik})$, and require matches on characteristics $Z_{ik}(C_i)$. For each risk set, we can estimate $S^{\text{IPCW}}_T(t | R_i)$. Inverse transform sampling from $S^{\text{IPCW}}_T(t)$ is then used to match candidate $i$ to a specific candidate $k \in R_i$. We can then impute patient $i$'s future status updates by copying over status updates from patient $k$. This procedure is repeated, until all candidates have a set of status updates ending with a waiting list removal (R) or death (D). 

\subsection{Modeling the expected remaining survival time for candidates}
\label{app:exp_rem_surv}
Here, we discuss the procedure used to estimate counterfactual survival curves, and how pseudo-outcomes are constructed for the log expected survival times. \\[0.2cm]
\subsubsection{IPCW survival curve estimate \& construction of pseudo-outcomes}
\noindent
\textbf{Propensity score model}\\
We are interested estimation of counterfactual waiting list survival, i.e. the probability that a patient is not yet delisted / has not yet died on the waiting list if transplantation were not available.  To correct for informative censoring by transplantation, we use a Cox model to predict the probability that a patient is censored over time, and estimate the counterfactual waiting list survival curve weighing observations by the inverse probability of being transplanted. Adjustment variables included in the Cox model for prediction of the transplant probability are candidate sex, candidate blood group, spline terms of candidate age, the disease group (congenital, polycystic, neoplasms, diabetes, glomerular disease, renovascular / vascular disease, tubular and interstitial disease, or other), the HLA-ABDR mismatch frequency (see section \ref{subsection:mmp}), the vPRA, the last reported PRA, and the candidate country of listing.\\[.3cm]
\par
\FloatBarrier
\noindent
\textbf{Construction of pseudo-outcomes}\\
As in \citeAC{tayobstatistical2017}, we are interested in directly modeling residual remaining survival time $T^{*}$, i.e. modelling
$$\mathbb{E}[\log(T^*)|Z] = \beta^\intercal Z.$$ A problem encountered in estimating such a model is that transplantation and censoring prevents Eurotransplant from observing the remaining survival times $T^*_{it}$ for most candidates. Tayob and Murray \citeAC{tayobstatistical2017} address this issue by constructing pseudo-observations for $\log(T_{ij}*)$ can be constructed, based on estimated counterfactual survival curve $\hat{S}^{IPCW}(t)$. We used this procedure to construct pseudo-observations for $\log(T_{it}*)$. Armed with pairs $(PO_{it}, Z_i)$, we can in principle estimate $\beta^{PO}$.

\subsubsection{Modelling the mean restricted survival time with Quasi-Least Squares}
With pairs $(PO_{it}, Z_i)$ we can model the expected log remaining survival time as
\begin{equation}
	\mathbb{E}[\log(T^*) | Z] = \beta^\intercal Z
	\label{eqn:model_tayob}
\end{equation}
However, in estimation of $\beta$ requires that we deal with correlations in $PO_{it}$ over $t$. Tayob and Murray faced a similar issue, where there is correlation between $T^*_{ij}$ for the different landmark times $j$, and addressed this by estimating $\beta$ with Generalized Estimating Equations (GEE) with an unstructured correlation matrix correlation over the landmark times $j$ (which requires $j(j+1)/2$ parameters). Unfortunately, this strategy is infeasible to us as potential outcomes $PO_{it}$ are indexed by continuous time $t$, i.e. all observed censoring times. Instead, we assume a Markov correlation structure for the potential outcomes, i.e. we assume that the correlation between measurements $PO_{is}$ and $PO_{it}$ decays with their separation in time: $$\texttt{Corr}(PO_{is}, PO_{it}) = \alpha^{|s-t|}.$$
Parameters $\alpha$ and $\beta$ to this model may be estimated with the \texttt{qlspack} R package, with Quasi-Least Squares \citeAC{Shults2014-di, xieqls2010}[p.94].
\par
For this model, we adjust for candidate age, candidate sex, whether the candidate has previously received a kidney transplantation, and the time the candidate has waited on dialysis at the moment of reporting the status update.

\subsection{The construction of risk sets of comparable candidates}
In the previous subsection, we discussed how the expected remaining survival time was modelled for candidates waiting on the Eurotransplant kidney waitling list. This information is used to construct, for each transplant recipient, a set of candidates who remain on the kidney waiting list who have comparable remaining survival times. Specifically, we required that $|\hat{\beta}^{\text{PO}}\ ^\intercal Z_k(C_{i}) - \hat{\beta}^{\text{PO}}\ ^\intercal Z_i(C_{i})| < 0.50$ to match transplant recipients to not-yet-transplanted waiting list candidates based on expected log 1-year-truncated waiting list survival.
\par 
Additional constraints were used to ensure that candidates contained within a risk set were comparable. For this, an adaptive procedure was implemented. This procedure always matched candidates on whether they were a re-transplantation candidate, and whether their last reported waiting list status was an active waiting list status. The procedure also tried to match candidates based on disease group, reason why they were non-transplantable, and candidate country of listing. However, these constraints were relaxed in case fewer than 50 candidates could be included in the risk set. Finally, the procedure also imposed constraints on the differences in accrued dialysis time and age at listing using pre-determined caliper widths.
\par 
In case matching according to all criteria fails to result in a risk set of sufficient size (that is, fewer than 50 candidates), we drop a discrete match criterion (first country, then reason non-transplantable, then disease group). In case dropping all discrete match criteria does not result in adequately sized risk set, we increase caliper widths for continuous variables.\\[.5cm]
\par
\noindent
\textbf{Example of a risk set}\\
In Table \ref{tab:example_match_table} we show an example of a constructed risk set for a female candidate who was transplanted after waiting for 4.5 years for a kidney transplantation with 6 years of accrued dialysis time. The first row of Table \ref{tab:example_match_table} shows that the transplanted patient is listed in 2014 in Germany for polycystic kidney disease. Remaining rows of Table \ref{tab:example_match_table} show 10 (out of 50) waiting list candidates in patient $i$'s risk set $R_i$. These patients remain at risk having waited 2162 days on dialysis for transplantation. They are also similar to the patient based on match characteristics: all matched candidates are patients with polycystic kidney disease waiting in Germany, and around age 60.
\begin{table}[h]
	\centering
	\caption{Example of the matched risk set for a selected recipient.}
	\resizebox{1\textwidth}{!}{
		
			\centering
	\begin{tabular}{rrllrrllll}
		\toprule
		Year & t & $C_i$ & $T_i$ & Days on dial & Age & ReTX & Urg. & Diag & Country\\
		\midrule
		\addlinespace[0.3em]
		\multicolumn{10}{l}{\textbf{To be imputed subject $i$}}\\
		\hspace{1em}2014 & 1602 & 1618 & - & 2162 & 64 & 0 & T & Polycystic & Germany\\
		\addlinespace[0.3em]
		\multicolumn{10}{l}{\textbf{Risk set $R_i$}}\\
		\hspace{1em}2017 & 1506 & 2229 & - & 1965 & 62 & 0 & T & Polycystic & Germany\\
		\hspace{1em}2010 & 1595 & - & 2346 & 2410 & 62 & 0 & T & Polycystic & Germany\\
		\hspace{1em}2012 & 1635 & - & 2837 & 2495 & 61 & 0 & T & Polycystic & Germany\\
		\hspace{1em}2014 & 1961 & - & 2619 & 1905 & 62 & 0 & T & Polycystic & Germany\\
		\hspace{1em}2016 & 1467 & - & 1747 & 2009 & 59 & 0 & T & Polycystic & Germany\\
		\hspace{1em}2015 & 1775 & 2255 & - & 2245 & 59 & 0 & T & Polycystic & Germany\\
		\hspace{1em}2016 & 1464 & - & - & 1869 & 58 & 0 & T & Polycystic & Germany\\
		\hspace{1em}2019 & 1567 & - & - & 2499 & 58 & 0 & T & Polycystic & Germany\\
		\hspace{1em}2010 & 1459 & 2333 & - & 1991 & 58 & 0 & T & Polycystic & Germany\\
		\hspace{1em}2011 & 1686 & 3531 & - & 1898 & 61 & 0 & T & Polycystic & Germany\\
		\bottomrule
\end{tabular}
	}

	\label{tab:example_match_table}
\end{table}
\FloatBarrier
With risk set $R_i$ and constructed IPCW weights, we can obtain a personalized estimate of the conditional probability of remaining waiting listed $t$ time units after $i$'s censoring time (i.e. $\hat{S}^{\text{IPCW}}_T(t|R_i, T > C_i)$). 

\subsection{Matching the patient to a particular patient in $R_i$}
We match the censored patient $i$ to a patient who remains at risk in the risk set $R_i$ with inverse transform sampling. This means that we (i) draw a random number $u$ from the uniform(0,1) distribution, and (ii) find the smallest matching $t$ such that $\hat{S}^{\text{IPCW}}_T(t|R_i, T > C_i) \leq u$). If such a $t$ exists, it corresponds to an event time for a patient $k \in R_i$. We can impute future status updates for patient $i$ by copying over the future status updates from patient $k$. 
\par
In case such a $t$ does not exist, the patient remains alive at least $\tau$ time units after the status update. We therefore need to match to a patient from the set of patients who remain alive $\tau$ days after $C_i$, i.e. $\{k \in R_i : T_{k} > C_i + \tau\}$. For this, we sampled patients with probability proportional to their inverse probability censoring weight, i.e. proportional to $\mathbb{P}[T_{k\tau} \leq C_{it} + 365 | T_{k\tau} > C_{it}].$

\FloatBarrier

\bibliographyAC{refs}

\section{Simulating post-transplant mortality and re-listing and re-transplantations}
\label{app:posttxp}
\setcounter{table}{0}

\bibliographystyleAC{myama}

\newcommand\Time{{\hspace{-.25pt}\intercal}}

\subsection{Post-transplant survival models \& relistings}
This section discusses how post-transplant survival and listing for a repeat transplantation is simulated based on patient and donor characteristics in the ETKidney simulator. For post-transplant survival, parametric survival models were used. The Kaplan-Meier estimator was used to determine the time-to-relisting relative to simulated post-transplant survival times. Both models were estimated on transplantations in the Eurotransplant region between 01-01-2011 and 01-01-2021. 

\subsubsection{Patient failure or repeat transplantation as the outcome}
Outcomes of interest after kidney transplantation are (i) patient survival, (ii) repeat transplantation, (iii) graft failure, and (iv) return-to-dialysis. For the prediction of post-transplant survival in the ETKidney simulator, we have chosen to define the event of interest as a kidney re-transplantation or patient death, whichever event occurs first. We have chosen not to model time-to-graft failure since centers in Eurotransplant do not work with standardized definitions of graft failure, and return-to-dialysis is not explicitly reported to Eurotransplant. For simulations, we assume that a patient would die on their event time, unless they are re-transplanted before the event materializes.

\subsubsection{Simulating patient failure date}
\label{subsection:sim_time_to_event}
Weibull accelerated failure time regression are used to simulate a time-to-event for patient death/re-transplantation. For this, we used the R \texttt{survival} library. The Weibull distribution in this library is parametrized with a shape parameter $k$ and scale parameter $\lambda = \beta^{\Time} x$, where $x$ are relevant patient and donor characteristics. The survival function for the Weibull distribution is given by:
$$S(t|x) = \exp\Bigg(-\Big(\frac{t}{\beta^{\Time}x}\Big)^{k}\Bigg).$$
After obtaining estimates for $\beta$ and $k$ based on historic data, we can simulate time-to-events for patient-donor pair $i$ by inverse transform sampling from this distribution. Specifically, we can draw a random number $u \sim \texttt{unif}(0,1)$ and simulate a patient's time-to-event as 
$$t = \hat{\mathbf{\beta}}^\Time \mathbf{x_i} \Big(-\log(u)^{\frac{1}{k}}\Big).$$
\par
For the default model supplied with the ETKidney simulator, we adjusted for donor characteristics (DBD/DCD, age, hypertension, last creatinine, diabetes, death cause, malignancy), patient characteristics (age, on dialysis, years on dialysis, repeat transplantation), and transplantation characteristics (year of transplantation, cross-country, standard or non-standard allocation, and the number of HLA-A, -B and -DR mismatches). We included shape parameters for the candidate's country of listing.

\subsubsection{Simulating time-to-relisting}
\label{subsection:sim_time_to_relist}
Most candidates who experience an (early) post-transplant event would list for a repeat transplantation. Such a time-to-relisting $r$ logically has to occur before a candidate's time-to-event $t$. To simulate relistings in the ETKidney simulator, we estimated the proportion of individuals that have relisted with Kaplan-Meier using $r/t$ as the time scale. In this, we stratify the curves by the candidate's time-to-event $t$ ($<$180 days, 180d-$<$1y, 1y-$<$2y, 2y-5y, 5y or more) and age at transplantation (0-17, 18-39, 40-49, 50-54, 55-59, 60-64, 65-69, 70-74, 75+). A relisting time $r$ can then be simulated by inverse transform sampling from the time-to-event $t$ and age-at-transplantation specific curve, i.e. (i) sample a random $u \sim \text{unif}(0,1)$, (ii) choose the first $s$ such that $\mathbb{P}[R_i/T_i > s] \geq u$, and (iii) calculating the time-to-relisting as $s*t$.
\par
\begin{figure}[h]
	\centering
	\includegraphics[width=\linewidth]{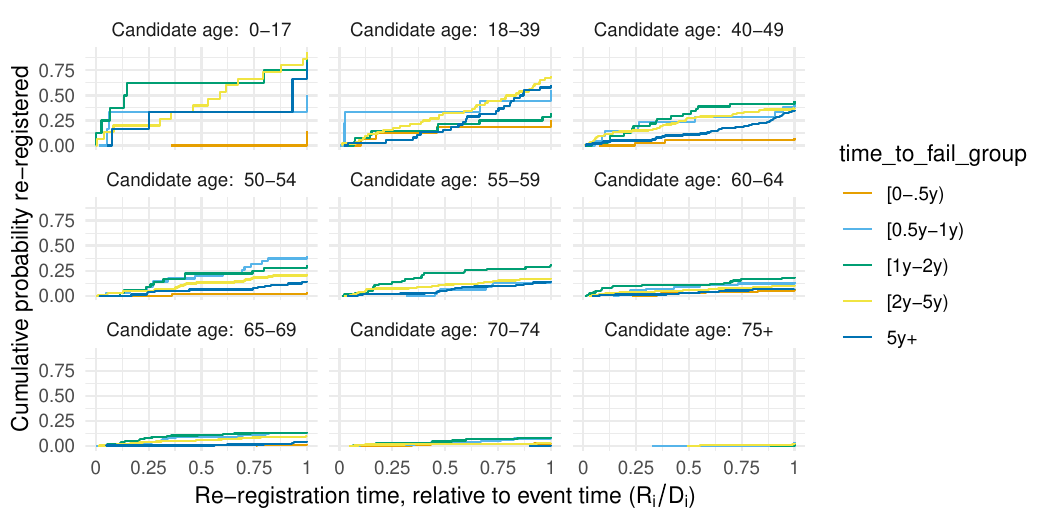}
	\caption{Kaplan-Meier curves used to simulate the time-to-relisting $R_i$, estimated on non-HU re-registrations between 2012 and 2020.}
	\label{fig:posttxp_surv}
\end{figure}

In case no such $t$ exists, the patient will not relist and therefore be a post-transplant death. Figure \ref{fig:posttxp_surv} shows estimates of the empirical distribution of $R_i/T_i$, stratified by groups of time-to-event. The figure clearly shows that the fraction of patients dying after transplantation without relisting depends on the failure time and the candidate's age. For instance, most pediatric candidates (0-17) are highly likely to list for repeat transplantation if they have an event within 5 years after listing, whereas very few older candidates (65+) list for repeat transplantation.
\FloatBarrier
\subsubsection{Constructing a synthetic re-registration}
\label{subsection:synth_registrations}
Subsections \ref{subsection:sim_time_to_event} and \ref{subsection:sim_time_to_relist} discussed how a time-to-event $t$ and time-to-relist $r$ are simulated in the ETKidney simulator at transplantation based on patient and donor characteristics. This is not sufficient for simulation on what would happen to this patient after transplantation in case the candidate lists for a repeat transplantation, because it is not known how the candidate's status would evolve when this candidate lists for a repeat transplantation. Here, we describe how synthetic re-listings are constructed in the ETKidney simulator by combining candidate information from the kidney recipient with status updates from an actual listing for a repeat transplantation.
\par 
Actual listings for a repeat transplantation have a known time-to-relisting $r_k$ and time-to-event $t_k = d_k + r_k$, where $d_k$ was potentially imputed using the status imputation procedure detailed in appendix \ref{app:imp_proc}. To find suitable status updates for a synthetic re-listing for candidate $i$, we can thus look for a relisting $k$ with $(r_k, t_k)$ similar to $(r_i, t_i)$. In practice, we could want to constrain matches further by also requiring $i$ and $k$ to match on other characteristics. The post-transplant module includes code to require such matches.
\par
Specifically, the post-transplant module by default finds a re-registration $k$ by:
\begin{enumerate}[noitemsep]
	\item matching $i$ and $k$ on candidate country of listing, and matching $i$ and $k$ on whether the candidate re-listed within 1 year of transplantation because such candidates are eligible for returned dialysis time (see \cite{manualKidney}). We also restrict matches based on continuous variables: the difference in age is at most 20 years, the difference in $r$ at most 2 years, the difference in $t$ is at most 1 year, and the difference in accrued dialysis time at most 3 years.	
	\item From matching registrations, selecting the $m=5$ re-registrations with the closest Mahalanobis distance between ($R_i$, $T_i$) and ($R_k$, $T_k$). 
	\item Sampling a random re-registration from the $m$ re-registrations.
\end{enumerate}
A synthetic re-registration is then constructed by combining patient attributes from patient $i$ with status updates from patient $k$. The post-transplant module only copies over urgency statuses. Other patient statuses are not copied over (allocation profiles, disease group updates, HLA updates, unacceptable antigens, and program choices), because these are patient-specific. For dialysis time, the initial dialysis time at re-listing is copied over from the synthetic re-listing, but updates to the dialysis time are not copied over. Unacceptable antigens are simulated based on the mismatched donor HLAs (see main text). Within simulations, synthetic re-listings thus do not change unacceptables, disease groups, allocation profiles, or their HLA.

\bibliographyAE{refs}

\newpage
\section{Supplementary Tables}
\label{app:supp}

\begin{table}[h]
	\caption{Input-output validation of the number of transplantations by vPRA, before and after introduction of the virtual crossmatch on 24-01-2023.}
	\resizebox{\linewidth}{!}{
	
\begin{tabular}{rrlrl}
\toprule
\multicolumn{1}{c}{ } & \multicolumn{2}{c}{Before virtual crossmatch} & \multicolumn{2}{c}{After virtual crossmatch} \\
\cmidrule(l{3pt}r{3pt}){2-3} \cmidrule(l{3pt}r{3pt}){4-5}
vPRA (\%) & simulated
(mean and 95\%-IQR) & actual & simulated
(mean and 95\%-IQR) & actual\\
\midrule
0\% & 2751 [2711-2792] & 2755 & \textbf{1460 [1428-1489]} & 1387\\
$>$0-84.9\% & 875 [842-909] & 861 & \textbf{446 [417-474]} & 494\\
85-94.9\% & 114 [99-128] & 117 & 59 [48-69] & 69\\
95+\% & 67 [56-78] & 75 & \textbf{34 [25-44]} & 46\\
\bottomrule
\end{tabular}

	}
	\label{tab:vpra_vxm}
\end{table}

\begin{table}[h]
	\caption{Input-output validation of the number of transplantations of ESP donors in candidates aged below 65 by country.}
	\resizebox{\linewidth}{!}{
		
\begin{tabular}{rrl}
	\toprule
	\multicolumn{1}{c}{ } & \multicolumn{2}{c}{Number of ESP transplantations in candidates aged below 65} \\
	\cmidrule(l{3pt}r{3pt}){2-3}
	Country of listing & simulated
	(mean and 95\%-IQR) & actual\\
	\midrule
	Austria & 19 [12-28] & 20\\
	Belgium & \textbf{14 [6-23]} & 3\\
	Croatia & 13 [7-21] & 8\\
	Germany & 156 [131-184] & 139\\
	Hungary & 3 [1-6] & 6\\
	\addlinespace
	Netherlands & \textbf{18 [8-30]} & 1\\
	Slovenia & \textbf{5 [2-10]} & 1\\
	\bottomrule
\end{tabular}

	}
	\label{tab:esp}
\end{table}

\end{document}